\documentclass[aps,prd,nofootinbib,twocolumn,floatfix]{revtex4}

\usepackage{amsmath}
\usepackage{amssymb}
\usepackage{color}
\usepackage{graphicx}

\def\aap{Astron. Astrophys.}
\def\aaps{Astron. Astrophys. Supplement}
\def\aj{Astron. J.}
\def\apj{Astrophys. J.}
\def\apjl{Astrophys. J. Lett.}
\def\apjs{Astrophys. J. Suppl. Ser.}
\def\araa{Annu. Rev. Astron. Astrophys.}
\def\jgr{J. Geophys. Res.}
\def\mnras{Mon. Not. R. Astron Soc.}
\def\nar{New Astron. Rev.}
\def\nat{Nature}
\def\pasa{Pub. Astron. Soc. Aust.}
\def\pasj{Publ. Astron. Soc. Jpn.}
\def\pasp{Publ. Astron. Soc. Pac.}
\def\prd{Phys. Rev. D}
\def\procspie{Proceedings of the SPIE}

\def\etal{\emph{et al.}}

\def\pm{\textrm{pm}}
\def\nm{\textrm{nm}}
\def\um{\mu\textrm{m}}
\def\mm{\textrm{mm}}
\def\cm{\textrm{cm}}
\def\meter{\textrm{m}}
\def\km{\textrm{km}}
\def\AU{\textrm{AU}}
\def\pc{\textrm{pc}}
\def\kpc{\textrm{kpc}}
\def\Mpc{\textrm{Mpc}}
\def\Gpc{\textrm{Gpc}}

\def\kHz{\textrm{kHz}}
\def\MHz{\textrm{MHz}}
\def\GHz{\textrm{GHz}}
\def\THz{\textrm{THz}}
\def\PHz{\textrm{PHz}}
\def\msec{\textrm{msec}}
\def\usec{\mu\textrm{sec}}
\def\arcmin{\textrm{arcmin}}
\def\arcsec{\textrm{arcsec}}
\def\mas{\textrm{mas}}
\def\uas{$\mu$\textrm{as}}
\def\nas{\textrm{nas}}
\def\pas{\textrm{pas}}
\def\fas{\textrm{fas}}
\def\ga{\gtrsim}
\def\la{\lesssim}
\newcommand{\mean}[1]{\ensuremath{\langle #1 \rangle}}
\newcommand{\InnerProduct}[2]{\ensuremath{\langle #1 | #2 \rangle}}

\begin{document}

\title{Arbitrary Transform Telescopes: The Generalization of Interferometry}
\author{Brian C. Lacki}\email{brianlacki@ias.edu} 
\affiliation{Institute for Advanced Study, School of Natural Sciences, Einstein Drive, Princeton NJ 08540, USA}
\date{January 2015}

\begin{abstract}
The basic principle of astronomical interferometry is to derive the angular distribution of radiation in the sky from the Fourier transform of the electric field on the ground.  What is so special about the Fourier transform?  Nothing, it turns out.  I consider the possibility of performing other transforms on the electric field with digital technology.  The Fractional Fourier Transform (FrFT) is useful for interpreting observations of sources that are close to the interferometer (in the atmosphere for radio interferometers).  Essentially, applying the FrFT focuses the array somewhere nearer than infinity.  Combined with the other Linear Canonical Transforms, any homogeneous linear optical system with thin elements can be instantiated.  The time variation of the electric field can also be decomposed into other bases besides the Fourier modes, which is especially useful for dispersed transients or quick pulses.  I discuss why the Fourier basis is so commonly used, and suggest it is partly because most astrophysical sources vary slowly in time.  
\end{abstract}

\maketitle

\section{Introduction}

Aperture synthesis interferometers are now a common instrument in astronomy that map sources on small angular scales.  Interferometers exploit the time/phase delay of a radio signal arriving at different receivers to measure these details \cite{Michelson20}. The typical understanding of how interferometers work is based on the Fourier transforms of the electric field -- first in time, then in space \cite{Tegmark09}.  Interferometers have been used for decades now, particularly in radio astronomy, where the long wavelengths lead to low angular resolution \cite{Kellermann01}.  

Normally, diffraction limits the resolution of a single-dish telescope with diameter $L$ observing radiation with wavelength $\lambda$ to $\Delta \theta \approx \lambda / L$.  But a pair of receivers in an interferometer measure the Fourier mode of the angular distribution of radiation on the sky with characteristic angles
\begin{equation}
\Delta \theta \approx \lambda / \ell,
\end{equation}
if $\ell$ is the separation of the receivers.  Each measured mode -- from each pair of receivers -- corresponds to a point on the $uv$ plane, which is basically the Fourier transform of the beam shape.  A single dish telescope can be viewed as a filled-in array of receivers that measures all of the Fourier modes within a disk of the $uv$ plane.  While a pair of interferometers provides only sparse information, more detailed geometrical information can be reconstructed by using an array of $\hat{N}$ interferometers (with ${\cal O}(\hat{N}^2)$ pairs), using Earth's rotation to rotate the $uv$ plane, or using a spread of different frequencies.

Although early interferometers actually channeled radiation from receivers to a central location (for example, through waveguides) to be correlated, with modern technology, the field amplitude at each receiver can simply be recorded digitally.  Then, the signals can be processed numerically later to reconstruct the image.  Digital processing of this kind is necessary for Very Long Baseline Interferometry (VLBI), where the receivers can be separated by thousands of kilometers \cite{Matveynko65Deller07}.  It is the method of choice for new or upgraded radio telescopes, including the Karl G. Jansky Very Large Array (JVLA) \cite{Perley09}, the Giant Metre-wave Radio Telescope (GMRT) \cite{Paciga11}, the Allen Telescope Array (ATA) \cite{Welch09}, and the Low Frequency Array (LOFAR) \cite{vanHaarlem13}.  Digital aperture synthesis is now enabling several arrays that are designed to detect 21 cm emission from the high-redshift intergalactic medium \cite{Furlanetto06}, among which are the Donald C. Backer Precision Array for Probing the Epoch of Reionization (PAPER) \cite{PAPER}, the Murchison Widefield Array (MWA) \cite{MWA}, the Australian SKA Pathfinder (ASKAP) \cite{Johnston08}, and MEERKat \cite{MEERKat}.  These are currently intended to be pathfinders for the proposed Square Kilometer Array (SKA), a global observatory intended to be built in the 2020s \cite{SKA}.

The Fast Fourier Transform Telescope (FFTT; \cite{Tegmark09}) is a refinement of digital interferometers.  The original proposal calls for a large $N \times N$ rectangular grid of simple dipole receivers ($\hat{N} = N^2$ total) with spacing $\ell \la \lambda$.  The gridding of the antennas allows the electric field to be transformed with the Fast Fourier Transform (FFT) in only ${\cal O} (\hat{N} \log \hat{N})$ operations \cite{Tegmark09,Tegmark10}.  The FFTT can map radio emission on angular scales ranging from the whole sky to $\lambda / (\ell N)$ radians.  It is thus ideal for measuring moderate to large brightness features over vast fields, the principle requirement for 21 cm reionization studies.  The array of dipole antennas at the Maip\'u Radio Astronomy Observatory was an analog version of the FFTT, although there were only 6 independent rows of antennas \cite{May84}.  The Waseda FFT Telescope is another early implementation of the FFTT, with $N = 16$ \cite{WasedaFFTT}.  The FFTT can be simplified into the fractal ``omniscope'' which is a nested series of grids while still providing resolutions over a wide range of scales \cite{Tegmark10}.  One of the low frequency pathfinders, the MIT Epoch of Reionization (MITEoR) telescope, is the first modern omniscope to be built \cite{Zheng14}.

But fundamentally, why do we use the Fourier transform in both time and space?  Cosines and sines are not the only possible basis for representing functions.  In fact, the Fourier transform is only one of a whole family of transformations, the Linear Canonical Transforms (LCTs).  These transforms do have physical meanings, which will be explored in this paper.  They can perform any area- and orientation-preserving linear mapping on space-frequency phase space: rotations, squeezing, and shearing.  Nor are we limited to LCTs: we could express the electric field as the linear sum of any set of basis functions, such as wavelets or pulses.  In principle, even more general non-linear transforms could be performed digitally.  

Understanding the meaning of these other transforms provides insight into how interferometry and beamforming works.  Although I have chosen to center my discussion on astronomical interferometers, the basic ideas apply whenever beamforming is used to receive or broadcast a signal.  Beamforming, of course, is the main principle behind phased array antennas, which are used in radar detection.  Nor is beamforming limited to electromagnetic radiation, but it also is used frequently with sound waves, such as sonar.

A note on notation: throughout this paper, I use $\delta(x)$ to refer to the Dirac delta function, $i$ refers to $\sqrt{-1}$, and an overline over a variable represents its complex conjugate.

\section{How standard interferometry works: an intuitive explanation \label{sec:InterferometryIntro}}

Interferometers exploit the phases of arriving electromagnetic waves to reconstruct the direction the radiation is coming from.  Consider polarized monochromatic electromagnetic waves emanating from a single direction ${\bf \hat{s}}$ on the sky.\footnote{The actual sum of the sky radiation can be described as a linear sum of such waves with different frequencies and coming from different directions.} The electric field at a point ${\bf r}$ and time $t$ is
\begin{equation}
{\bf E} = {\bf E_0} \exp\left[\frac{2 \pi i}{\lambda} ({\bf \hat{s}} \cdot {\bf r} + c t)\right]
\end{equation}
if the wavelength is $\lambda$.  I will consider the case of a 1D array, but the basic ideas can easily be generalized to a 2D array.  In addition, I will suppose that all of the receivers are located on the ground ($z = 0$).  The receivers in a 1D array are located at different positions in the $x$-direction.  The electric field on the ground is
\begin{equation}
{\bf E}(x) = {\bf E_0} \exp\left[\frac{2 \pi i}{\lambda} (x \sin \theta + c t)\right]
\end{equation}
where $\theta$ is the zenith angle of the source.

\begin{figure*}
\includegraphics[width=6cm]{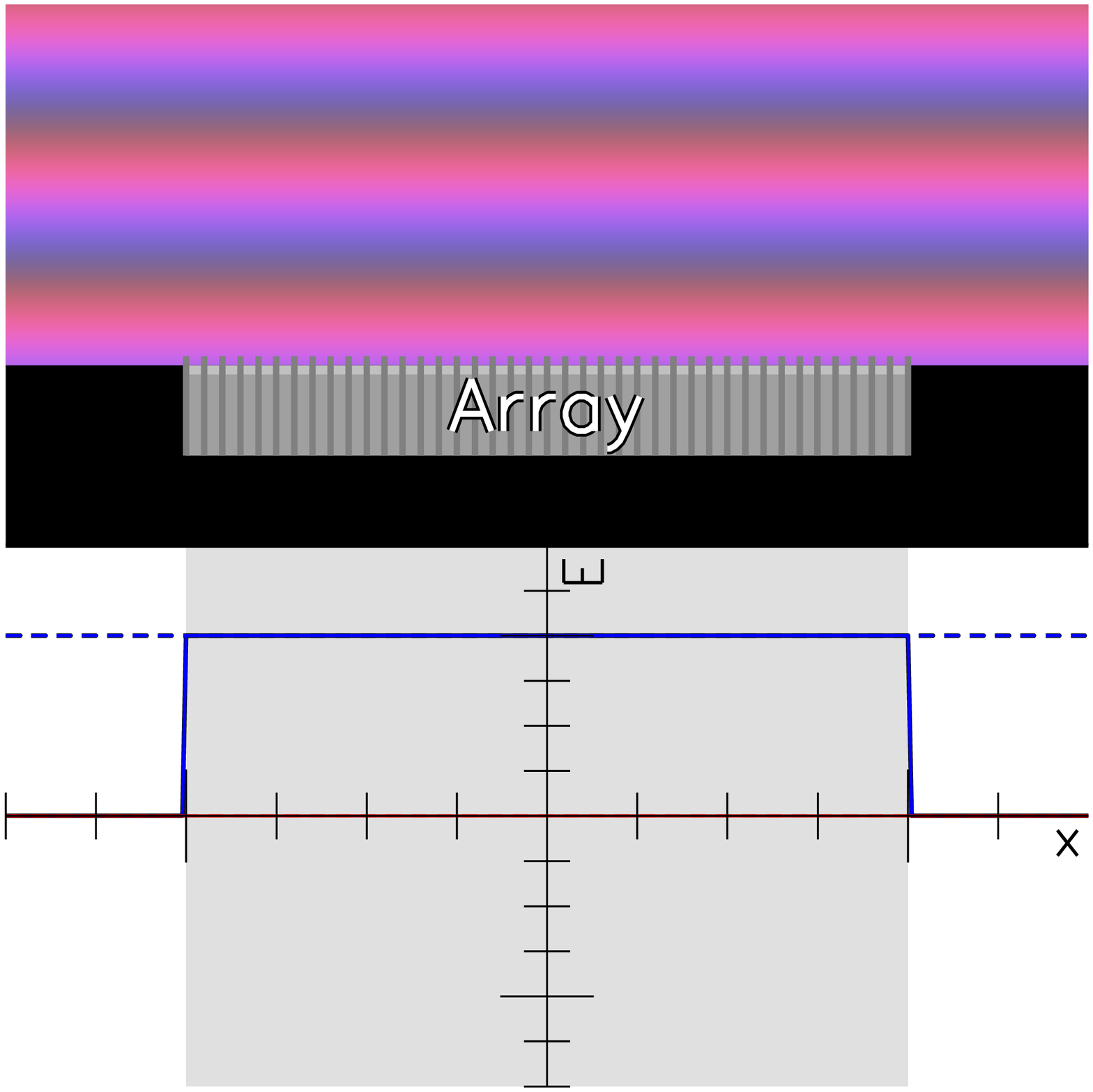}\,\includegraphics[width=6cm]{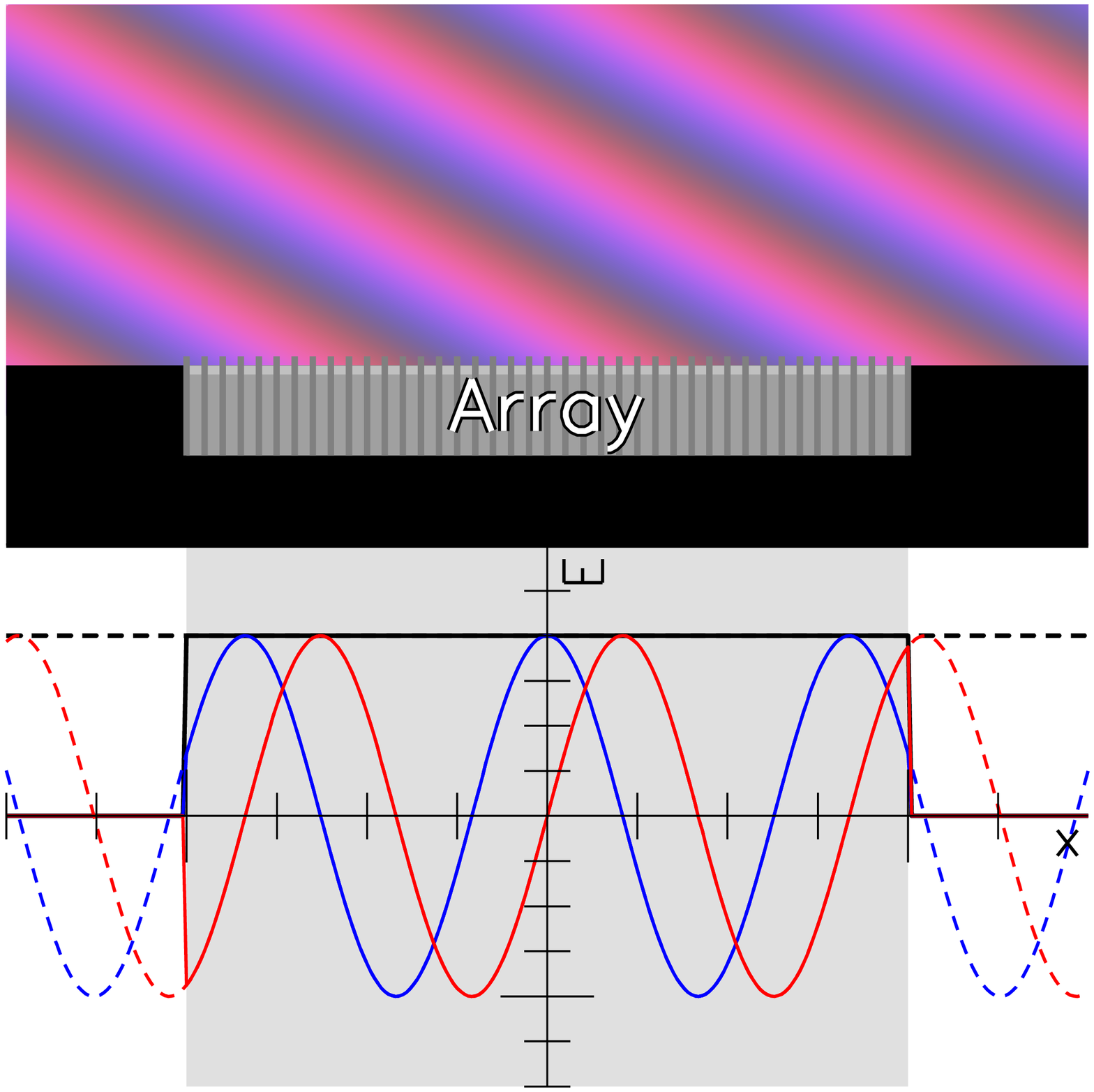}\,\includegraphics[width=6cm]{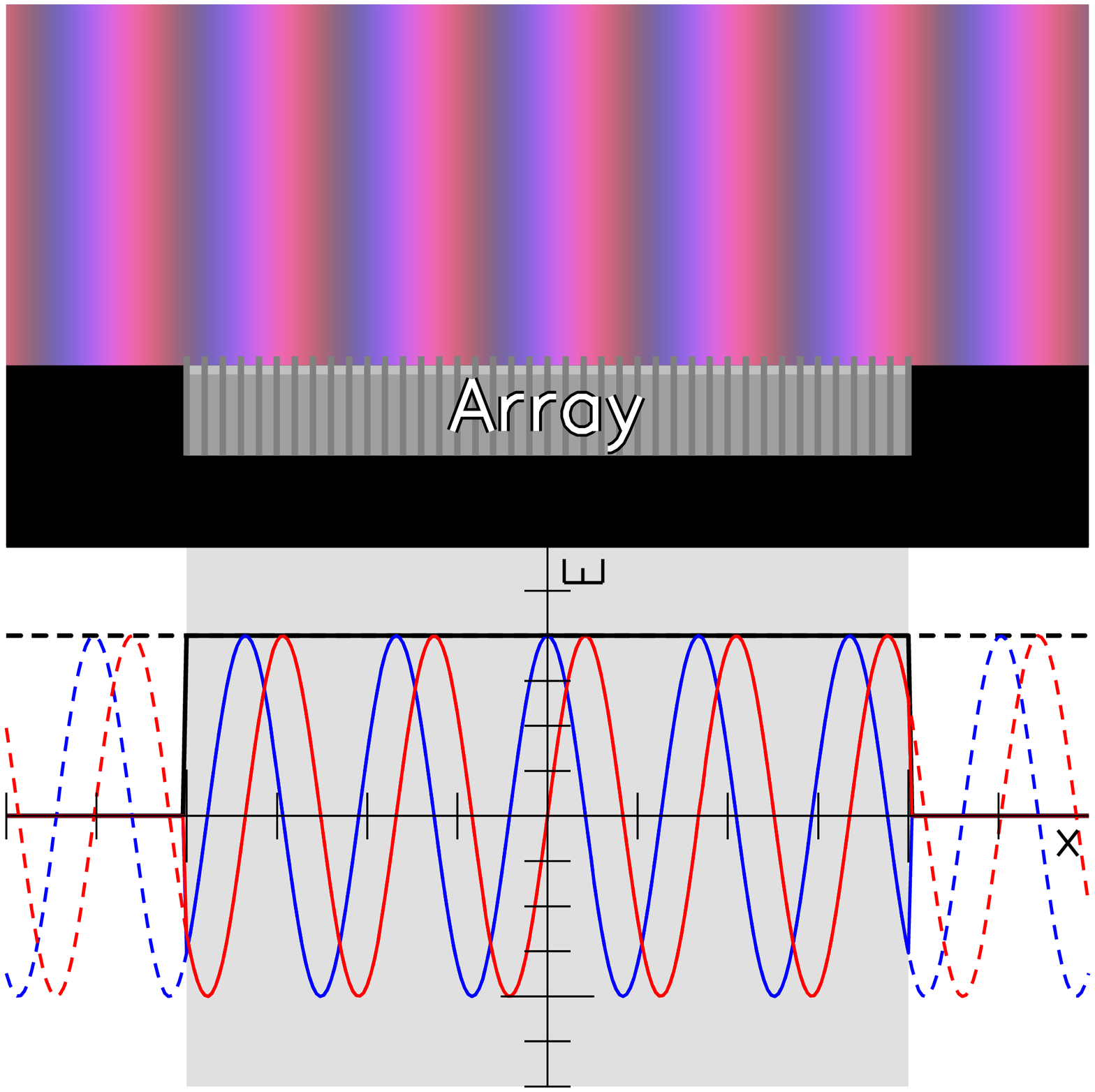}
\caption{As the angular altitude of a source of plane waves increases, the projected spatial wavelength of the plane waves decreases due to projection effects.  The components of the electric field are imaginary (red) and real (blue), with a magnitude shown in black.  The solid lines include only the signal that lands on the array, while the dotted lines show how the signal continues beyond its bounds.\label{fig:Interferometry}}
\end{figure*}

If the source is at the zenith, the phase of the electric field is the same everywhere on the ground (Figure~\ref{fig:Interferometry}).  However, as the zenith angle increases, it takes radiation longer to reach one side of the array than the other because the path lengths are slightly different.  The electric field on the ground at any instant therefore oscillates on the $x$-axis, with spatial wavelength $\lambda_x = \lambda / \sin \theta$ (Figure~\ref{fig:Interferometry}).  The spatial Fourier transform of this electric field is particularly simple:
\begin{equation}
\label{eqn:FourierE}
{\cal F}[{\bf E}](\kappa) = {\bf E_0} \times \delta\left(\kappa - \frac{\sin \theta}{\lambda} \right) = {\bf E_0} \times \delta\left(\kappa - \frac{1}{\lambda_x} \right).
\end{equation}
In Fourier space, all of the power is concentrated at an inverse wavelength $\kappa$ with a simple dependence on $\theta$.\footnote{I use $\kappa = 1/\lambda$ instead of the wavenumber $k = \kappa / (2\pi)$ throughout the paper.  This avoids the need to divide by $2 \pi$ when considering the units, although the equations themselves then do use a factor of $2 \pi$.}  Thus, the Fourier transform of the field describes the direction the radiation is coming from (bottom panels of Figure~\ref{fig:Interferometry}).  Since the Fourier transform is a linear operator, additional sources sum linearly in the Fourier transform of the electric field.  The interferometer can be ``aimed'' at a given direction ${\bf \hat{s}_0}$ by adding a time delay
\begin{equation}
\Delta t = {\bf \hat{s}_0} \cdot {\bf r} / c
\end{equation}
to each receiver, adjusting the relative phases of the electric field at each position.

When considering radiation of many different wavelengths, there appears to be an ambiguity in equation~\ref{eqn:FourierE}.  How can we tell whether a short wavelength mode is a high frequency wave coming from near the horizon or a low frequency wave coming from near the zenith?  This is easily resolved by examining the signal in time domain.  The time frequency of the signal is the same everywhere on the ground plane; it is merely delayed at some points more than others.  Interferometry works because the projection of the wave onto a plane stretches it out in spatial wavelength without changing its time frequency.

Each pair of receivers $a$ and $b$ is separated by a distance $(\Delta x_{ab}, \Delta y_{ab})$; the separation can also be written in terms of the $uv$ coordinates:
\begin{equation}
(u_{ab},v_{ab}) = \left(\frac{\Delta x_{ab} \sin \theta_{\rm source}}{\lambda}, \frac{\Delta y_{ab} \sin \theta_{\rm source}}{\lambda}\right).
\end{equation}
The pair measures a single Fourier mode of the electric field with wavelength $(u_{ab} \lambda, v_{ab} \lambda)$, which only provides information about the distribution of radiation on the sky on an angular scale of
\begin{equation}
(\theta_u,\theta_v)_{ab} = (1/u_{ab}, 1/v_{ab}).
\end{equation}
Thus, traditional interferometers are insensitive to structures with angular scales greater than $(\min(1/u_{ab}),\min(1/v_{ab}))$ or less than $(\max(1/u_{ab}),\max(1/v_{ab}))$.  An important goal of interferometers is to ``fill in'' as much of the $uv$ plane as possible, and this can be aided by allowing the Earth's spin to rotate the sky and by using different frequencies.  

The FFTT is the simplest interferometer to understand without obscuring the basic principle.  In a standard FFTT, the receivers are arranged on a $N \times N$ rectangular grid with separation $\Delta l < \lambda_{\rm min}$.  By Nyquist's theorem, the electric field everywhere on the grid can be reconstructed with a spatial resolution $\lambda_{\rm min} / 2$.  A Fast Fourier Transform can be directly be performed on the electric fields measured at each receiver to produce an electric field image, which is then squared to give the intensity image \cite{Tegmark09}.  Structures on angular scales ranging from $\lambda_{\rm min} / (N \lambda)$ to the whole sky can be mapped this way with the FFTT.

The spatial Fourier Transform works because \emph{it measures the projection effect} on the time-domain Fourier modes.

\subsection{Complications in real interferometers}
\label{sec:InterferometerComplications}

Of course, it is often impractical to measure the electric field everywhere as a continuous function.  The subtleties in understanding interferometry (particularly the use of ``visibilities'')  arise because of the limited sampling of the electric field.  

Actual interferometers are composed of a series of discrete elements, which could be anything from simple dipole antennas to single dish telescopes.  These receivers each have a primary beam $A({\bf \kappa})$ that describes their sensitivity to radiation that comes from different directions ${\bf \kappa}$ on the sky.  A large single dish's primary beam is typically confined to a narrow region in the sky with radius $\lambda / D_{\rm dish}$ for a dish diameter $D_{\rm dish}$.  Dipole antennas have sensitivity over much of the sky, which is why they are proposed for FFTTs.  The measured electric field (or intensity) is the sky intensity convolved with $\sqrt{A({\bf \kappa})}$ (or $A({\bf \kappa})$).  I assume that the antennas all have the same response, are all pointed in the same direction, and there is only one point source in the sky, so that the antenna response can be approximated as a constant $\sqrt{A}$ (or $A$ in intensity).

The other limitation, mainly faced by traditional interferometers, is that the $uv$ plane may be sparsely sampled.  Interferometers, whether FFTTs or traditional, are insensitive to the electric fields outside of their collective area.  In addition, sparse interferometers only have a few measurements of the electric field from which to extrapolate the entire $uv$ plane.  Thus, the measured electric fields are filtered by the interferometer's sampling on the ground, $S({\bf x})$:
\begin{equation}
\label{eqn:SFilteredE}
E_{\Pi}^{\prime} ({\bf x}) = \sqrt{A} E_{\Pi} ({\bf x}) \times S({\bf x}),
\end{equation}
for a polarization $\Pi$.  Then the Fourier transform of the electric field is the convolution of its Fourier modes with the dirty beam $P_{\rm dirty} = {\cal F} [S]$:
\begin{equation}
E^{\rm dirty}_{\Pi} ({\bf \kappa}) = {\cal F}[E^{\prime}_{\Pi}] ({\bf \kappa}) = \sqrt{A} {\cal F} [E_{\Pi}({\bf x})] \ast P^{\rm dirty}.
\end{equation}
The result is the dirty map.  Some method is then necessary to deconvolve the dirty map with the dirty beam to get the sky distribution of radiation.   

In traditional interferometry, these difficulties are dealt with by calculating visibilities for each pair of receivers,
\begin{equation}
V (u_{12}, v_{12}) = \mean{E_1 E_2},
\end{equation}
averaged over time.\footnote{Note that intensity (surface brightness) has units of the electric field \emph{squared}.}  The van Cittert-Zernike theorem relates the visibilities to the Fourier transform of the sky intensity (e.g., \cite{Wilson09}).  Specifically, the dirty map is calculated as
\begin{equation}
I_{\rm dirty} ({\bf \kappa}) = A {\cal F} [V] \ast P_{\rm dirty}.
\end{equation}
Because of the sparsity of traditional interferometers, their dirty beams may be very complicated.  Algorithms like CLEAN \cite{Hogbom74} deconvolve the dirty map with the dirty beam.

\section{Digital optics through Linear Canonical Transforms \label{sec:DigitalLCTs}}

\subsection{A whole family of integral transforms \label{sec:LCTs}}

The Fourier transform is actually just one of the family of linear canonical transforms (LCTs).  The LCTs are the set of all linear integral transforms that conserve volume and orientation in position-frequency space.  They can be represented by the ABCD matrix 
\begin{equation}
{\bf M} = \left( {A \atop C}\ {B \atop D} \right)
\end{equation}
with a determinant of 1 \cite{Bernardo96,Stern06}.  The LCT synthesizes a signal at a new position coordinate $x^{\prime}$ and a new frequency (inverse wavelength) coordinate $\kappa^{\prime}$ out of the old position $x$ and frequency $\kappa$, as given by
\begin{equation}
\left({x^{\prime} \atop \kappa^{\prime}}\right) = {\bf M} \left({x \atop \kappa}\right).
\end{equation}
There is no unique definition for an LCT, but the following is usually used \cite{Ozaktas01}:
\begin{equation}
[{\bf M}f(x)](x^{\prime}) = \left\{ \begin{array}{r} \displaystyle \sqrt{\frac{-i}{B}} \int_{-\infty}^{\infty} e^{\pi i (A x^2 - 2 x x^{\prime} + D x^{\prime 2})/B} f(x) dx \\
(B \ne 0) \\ \\
\sqrt{D} e^{\pi i C D x^{\prime 2}} f(D x) \mspace{100mu} (B = 0) \end{array} \right.
\end{equation}

The Wigner distribution provides a geometric analogy for the effects of LCTs.  It measures how much of the signal is at each position and frequency \cite{Almeida94,Bultheel06}:
\begin{equation}
[{\cal W}f](x,\kappa) = \frac{1}{\sqrt{2 \pi}} \int_{-\infty}^{\infty} f\left(x + \frac{u}{2}\right) \overline{f\left(x - \frac{u}{2}\right)} e^{-2\pi i \kappa u} du.
\end{equation}
If $f(x)$ is a delta function, then its Wigner distribution is a vertical line (taking $x$ to be horizontal and $\kappa$ as vertical).  The function $\exp(2 \pi i \kappa x)$ is a horizontal line in Wigner space.  Diagonal lines in Wigner space are linear chirps.

A form of Parseval's theorem applies to LCTs,
\begin{equation}
\int_{\infty}^{\infty} |{\bf M}f(x)|^2 = \int_{\infty}^{\infty} |f(x)|^2.
\end{equation}
As such, an LCT cannot increase the total ``energy'' in a signal.  What they \emph{can} do is carve up Wigner space into slices that are aligned with a signal's energy.  This concentrates the energy into a small bin along one direction, making the signal easier to detect.

The LCTs are simply the composition of a few basic operations in Wigner (position-frequency) space \cite{Bultheel06}.  The first basic operation is rotation through some angle $\alpha$:
\begin{equation}
{\cal F}_{\alpha} = \left({\cos \alpha \atop -\sin \alpha}\ {\sin \alpha \atop \cos \alpha}\right).
\end{equation}
Rotation mixes the position domain with frequency domain.  When $\alpha = \pi / 2$, the LCT is simply the Fourier transform:
\begin{equation}
{\cal F} = \left({0 \atop -1}\ {1 \atop 0}\right).
\end{equation}
Then the new position coordinate is $\kappa$.  For other angles, the rotation is a \emph{Fractional Fourier Transform} (FrFTs).  FrFTs use chirps as their orthogonal basis functions.  The second basic operation is shear by some quantity $q$:
\begin{equation}
{\rm Fr}_q = \left({1 \atop 0}\ {q \atop 1}\right)
\end{equation}
The shear of position/frequency space is also known as a \emph{Fresnel transform}.  The final basic operation is magnification by a factor $\mu$, or
\begin{equation}
{\cal M}_{\mu} = \left({\mu \atop 0}\ {0 \atop 1/\mu}\right).
\end{equation}

Nor are LCTs limited to real numbers.  The Laplace transform is well known to be the Fourier transform of the imaginary frequencies, and is given by:
\begin{equation}
{\cal L} = \left({0 \atop i}\ {i \atop 0}\right).
\end{equation}
Of course, other transforms can be constructed out of these basic operations.

\subsection{\label{sec:MatrixOptics} LCTs and matrix optics: The basic optical elements}

The LCTs are deeply connected with optics.  Quadratic phase systems can be represented by an $ABCD$ matrix that corresponds directly to the $ABCD$ matrix of an LCT \cite{Ozaktas01}.  Suppose the electric field of a light wave at location 1 is described in terms of a position $x_1$ and spatial frequency $\kappa_1$.  The light wave is processed by some optical system ${\bf M}$ and reaches location 2, where the electric field is described in terms of position $x_2$ and spatial frequency $\kappa_2$.  Then the optical system's effects can be represented as
\begin{equation}
\left({x_2 \atop \kappa_2}\right) = {\bf M}\left({x_1 \atop \kappa_1}\right).
\end{equation}
For monochromatic light, the spatial frequency directly corresponds to a ray's angle with the optical axis.

${\bf M}$ is a composition of the $ABCD$ matrices of several basic elements.  If a light wave moves a distance $d$ through free space, the matrix is a shear matrix
\begin{equation}
{\bf M} = \left({1 \atop 0}\ {d \lambda \atop 1}\right)
\end{equation}
This is of course the Fresnel transform, named for Fresnel diffraction.  A thin lens with focal length $f$ shears in the frequency axis instead:
\begin{equation}
{\bf M} = \left({1 \atop -1/(\lambda f)}\ {0 \atop 1}\right)
\end{equation}
Any LCT can be constructed out of these elements \cite{Ozaktas01,Moreno05}.  

With digital optics, there are none of the usual physical limits on what material properties are realistic or not.  One can easily emulate a material with index of refraction that is less than 1, negative, or even complex.  Digital optics that mimic negative index of refraction materials can effectively act as superlenses, useful for near-field imaging of sources closer than one wavelength (see section~\ref{sec:Evanescence}).

\subsection{Lenses as Fourier transformers \label{sec:FourierLenses}}

As is well known, a lens performs the Fourier transform of the electric field.  We can confirm this with the $ABCD$ formalism.  Consider a lens with focal length $f$.  It focuses light from infinity into an image on a plane a distance $f$ after the lens.  Take the input field to be located a distance $f$ in front of the lens and the output field to be $f$ behind the lens.  The $ABCD$ matrix is
\begin{align}
{\bf M}_{\rm focus} & = \left({1 \atop 0}\ {\lambda f \atop 1}\right) \left({1 \atop -1/(\lambda f)}\ {0 \atop 1}\right) \left({1 \atop 0}\ {\lambda f \atop 1}\right).\\
                    & = \left({0 \atop -1/(\lambda f)}\ {\lambda f \atop 0}\right) 
\end{align}  
This is just the Fourier transform LCT matrix done in units of $\sqrt{\lambda f}$.  That is, we perform a ``magnification'' by a factor $1/\sqrt{\lambda f}$ on the input plane to get a unitless coordinate system, perform a Fourier transform, and then perform a ``magnification'' by a factor $\sqrt{\lambda f}$ on the output to get it back into physical units.

\section{Fractional Fourier Transforms: How to focus an interferometer \label{sec:FracInterferometer}}

Fractional Fourier Transforms lie in between the standard first-order Fourier transform and a zeroth-order Fourier transform -- the identity transform.  One can think of the identity and Fourier transforms as two limiting extremes.  If the source is so close that it lies within the array, it has a position but no single direction in the sky.  The identity transform on the electric field is then most useful.  If the source is at infinity, it lies in a definite direction, but it has no ``position'' within the array.  Then the first-order Fourier transform of the electric field is necessary to recover its direction.  This thought experiment suggests that if a source is at intermediate distance, its position and direction are both moderately defined, and one would use some intermediate fractional-order Fourier transform to find it.  This turns out to be the case.

\subsection{Parallax and FrFTs}
\begin{figure}
\includegraphics[width=8cm]{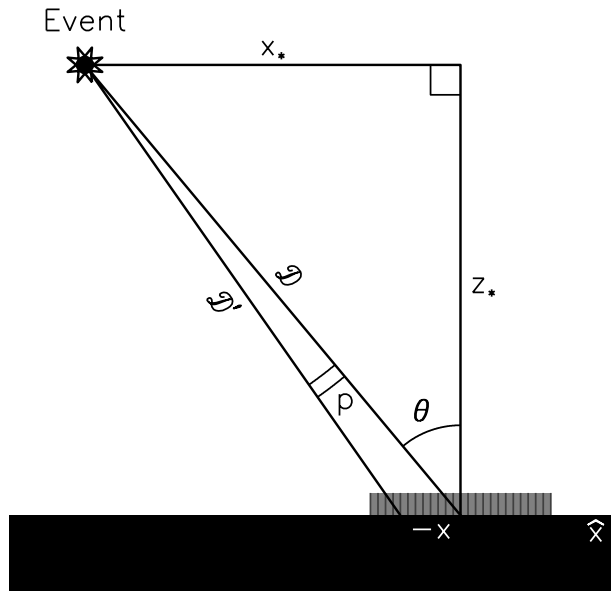}
\caption{The geometry of the fractional interferometer.\label{fig:FrFTGeometry}}
\end{figure}

\begin{figure}
\includegraphics[width=8cm]{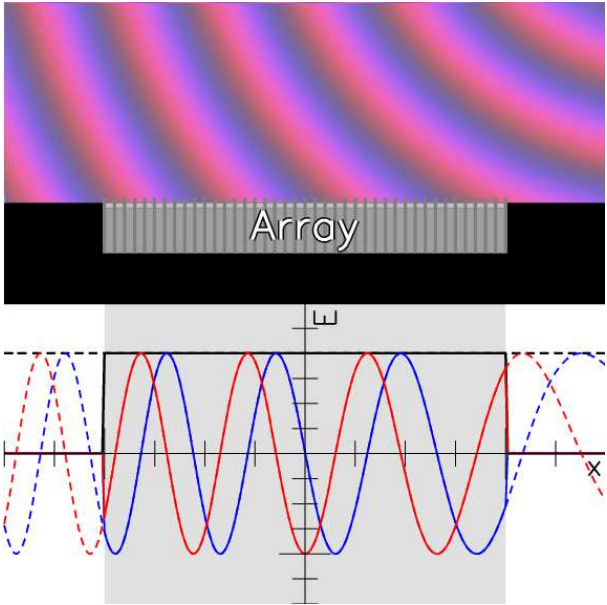}
\caption{The spatial frequency of waves from a nearby source changes with distance.  The signal on the ground is then a chirp.  This smears out the image in standard interferometers.  The lines styles are the same as in Figure~\ref{fig:Interferometry}. \label{fig:FractionalInterferometry}}
\end{figure}

If the source is merely very distant but not at infinity, the source displays parallax between different parts of the interferometer (Figure~\ref{fig:FrFTGeometry}).  Because of parallax, the radiation \emph{cannot} be effectively described as coming from a single direction, even if the source is localized (Figure~\ref{fig:FractionalInterferometry}).  The spatial wavelength of the radiation is smeared out, and the Fourier transform normally implemented by interferometers performs poorly.  

This can be seen in detail by reconsidering the case of a 1D array, but now with a point source located at a position $x = -x_{\star}$ and $z = z_{\star}$ (Figure~\ref{fig:FrFTGeometry}).  Its distance from the center of the array is ${\cal D}$.  The electric field at any given time is:
\begin{multline}
{\bf E}(x,z,t) =  \frac{{\bf E_0}}{\sqrt{(x + x_{\star})^2 + (z - z_{\star})^2}} \\
\times \exp\left[\frac{2 \pi i}{\lambda} (\sqrt{(x + x_{\star})^2 + (z - z_{\star})^2} + c t)\right].
\end{multline}
The electric field on the ground has a complicated dependence on position:
\begin{multline}
{\bf E}(x) = \frac{{\bf E_0}}{\sqrt{(x + x_{\star})^2 + z_{\star}^2}} \exp\left(\frac{2 \pi i c t}{\lambda}\right)\\
\times \exp\left[\frac{2 \pi i}{\lambda} \sqrt{(x + x_{\star})^2 + z_{\star}^2}\right],
\end{multline}
and thus cannot be inverted easily.  The spatially oscillating part of ${\bf E}(x)$ is an exponential of the form $\exp(i \phi)$.  The phase $\phi$ is
\begin{align}
\label{eqn:PhiTaylor}
\phi(x) & = \frac{2 \pi}{\lambda} {\cal D}^{\prime} \approx \frac{2 \pi {\cal D}}{\lambda} \left[1 + \left(\frac{x}{{\cal D}}\right) \sin \theta + \frac{x^2}{2{\cal D}^2} \cos^2 \theta\right]
\end{align}
when considering regions where $x \ll {\cal D}$.  

Taking a spatial derivative of $\phi$ amounts to measuring the instantaneous spatial frequency,
\begin{equation}
2\pi \kappa = \frac{d\phi}{dx} \approx \frac{2 \pi}{\lambda} \left(\sin \theta + \frac{x}{{\cal D}} \cos^2 \theta\right).
\end{equation}
This frequency is what interferometers usually measure, and the first term is just the projection effect described in Section~\ref{sec:InterferometryIntro}.  But if ${\cal D}$ is not infinite, then the spatial frequency does depend on position and the electric field can be approximated as a linear chirp \cite{Almeida94}.  The frequency slide is
\begin{equation}
2\pi \frac{d\kappa}{dx} = \frac{d^2 \phi}{dx^2} \approx \frac{2 \pi}{\lambda} \frac{1}{D} \cos^2 \theta.
\end{equation}

What we need is a transform that breaks the signal into chirps, measuring the power in a domain intermediate between position and frequency.  The Fractional Fourier Transform \cite{Namias80,McBride87},
\begin{multline}
\label{eqn:FrFT}
{\cal F}_{\alpha}[f(\chi)](\chi^{\prime})  = \sqrt{1 - i \cot \alpha}\ \times \\
\int_{-\infty}^{\infty} \exp[\pi i (\cot \alpha\ \chi^{\prime 2} - 2 \csc \alpha\ \chi^{\prime} x + \cot \alpha\ \chi^2)] f(\chi) d\chi.
\end{multline}
is suited for this purpose.  (I use $\chi$ and $\chi^{\prime}$ here to emphasize that the position and direction variables are \emph{unitless}.)  The angle $\alpha$ can be interpreted as the order $Q \equiv \alpha (2/\pi)$ of the FrFT.  The identity operation has $Q = 0$ ($\alpha = 0$), while the standard Fourier transform has $Q = 1$ ($\alpha = \pi / 2$).  With my conventions, $x^{\prime}$ reduces to $\kappa$ for $Q = 1$.  Although FrFTs are not (directly) used in astronomical interferometry to my knowledge, they were defined in 1980 with awareness spreading in fields such as optics (e.g., \cite{FrFTApplications}), signal processing \cite{Almeida94}, and quantum mechanics (e.g., \cite{Namias80}) through the 1990s and 2000s.  

\begin{figure}
\includegraphics[width=8cm]{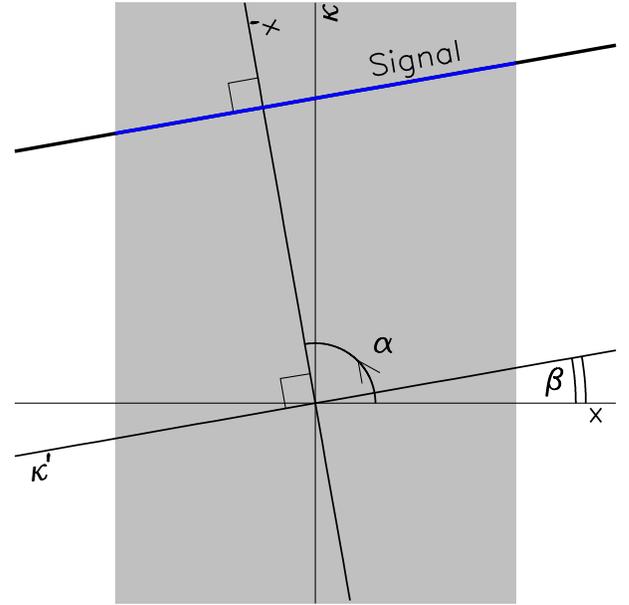}
\caption{A plot of the distribution of a chirp's signal in Wigner space (thick black/blue line).  An interferometer can focus on a source by rotating by an angle $\alpha$ so that the energy is all at one position $x^{\prime}$.  The coordinates are in natural units of $\lambda$ for $x$ and $1/\lambda$ for $\kappa$.  \label{fig:WignerFrFT}}
\end{figure}

In Wigner space, the distribution of the energy in ${\bf E}(x)$ is a diagonal line as shown in Figure~\ref{fig:WignerFrFT}.  The FrFT rotates position-frequency space so that the line of energy becomes vertical -- a $\delta$-function (Section~\ref{sec:LCTs}).  Then one can just read off the chirp's frequency derivative and use it to infer the distance of the source.  But we must choose the ``angle'' $\alpha$ correctly in order to optimally concentrate the chirp's power.  

The frequency $\kappa$ plotted in Figure~\ref{fig:WignerFrFT} is $1/(2\pi) \times (d\phi/dx)$.  Say the units for position are $s$ and the units of $\kappa$ are then $1/s$, in order to put everything into a unitless form (with $\chi = x/s$).  The slope of the chirp in Wigner space is
\begin{equation}
\tan \beta = \frac{d(\kappa s)}{d(x/s)} = \frac{s^2}{2 \pi} \frac{d^2 \phi}{dx^2} \approx \frac{s^2}{\lambda {\cal D}} \cos^2 \theta
\end{equation}
From this I infer that $\beta = \tan^{-1} [-s^2/(\lambda D) \cos^2 \theta]$, and the necessary rotation angle is
\begin{equation}
\label{eqn:alphaFocus}
\alpha = \frac{\pi}{2} + \beta \approx \frac{\pi}{2} + \frac{s^2}{\lambda {\cal D}} \cos^2 \theta.
\end{equation}
In other words, we must perform a Fractional Fourier Transform of order $Q \approx 1 + (2/\pi) (s \cos \theta)^2 / (\lambda {\cal D})$.  I call an interferometer that performs a FrFT a \emph{fractional interferometer}.

In astronomy, the sources are typically \emph{very} distant and the units are fairly small, so that $Q \to 1$.  The most ``natural'' units are the wavelength of the radiation $\lambda$, in which case the deviation of $Q$ from unity is a mere $\sim \lambda / {\cal D} \ll 1$.  Then $x^{\prime}$ is very nearly $\kappa$ and has practically the same units of inverse wavelength.  In this limit, $\cot \alpha \to -\beta + {\cal O}(\beta^3)$ and $\csc \alpha \to 1 + {\cal O}(\beta^2)$.  The calculation of the FrFT is easy in this limit:
\begin{equation}
\label{eqn:FrFTQ1}
{\cal F}_{\alpha}[E](\chi^{\prime}) \approx \exp[-\pi i \beta \chi^{\prime 2}]\ {\cal F}[E(\chi)\ \exp(-\pi i \beta \chi^2)],
\end{equation}
or after restoring units,
\begin{multline}
{\cal F}_{\alpha}[E](x^{\prime}) \approx \exp\left[-\pi i \frac{s^4 x^{\prime\,2}}{\lambda^2 D^2} \cos^2 \theta\right]\ \times\\
{\cal F}\left[E(x)\ \exp\left(-\pi i \frac{x^2}{\lambda D} \cos^2 \theta\right)\right].
\end{multline}
These approximations apply under the loose condition that $\beta^2 \chi \chi^{\prime} \ll 1$.  Since the largest $x$ in the problem is generally the array size $L$, and the largest $x^{\prime}$ is related to the minimum separation of the antennas ($x^{\prime} \la 1/\ell$), this is true if $L \ll \ell (\lambda D)^2/s^4$.  If one works in a ``natural'' unit system suggested above, with $s = \lambda$, then the condition is almost trivially true, $L \ll \ell (D/\lambda)^2$.

Applying a FrFT instead of a FFT therefore basically amounts to a chirp multiplication before and after the Fourier transform.

\subsection{The Fractional Fourier Transform as focus}
Just as a lens performs an ordinary Fourier transform when it focuses light from infinity (section~\ref{sec:FourierLenses}), it performs a FrFT when it focuses light from distance ${\cal D}$.  Specifically, if we again consider a lens with focal length $f$, then the electric field in an output plane a distance $d$ behind the lens is a Fractional Fourier transform of the field in an input plane a distance $d$ in front of the lens \cite{Bernardo94}.  This can be seen with a matrix optics analysis.  The transfer matrix for light propagating distance $d$, passing through a lens with focal length $f$, and propagating $d$ again is \cite{Bernardo96,Moreno05}
\begin{equation}
{\bf M}_{\rm focus} = \left({1 - d/f \atop -1/(\lambda f)}\ {d\lambda (2 - d/f) \atop 1 - d/f}\right).
\end{equation}
A transfer matrix of this type can be converted into a FrFT matrix by a scaling with the correct units:
\begin{align}
\nonumber {\cal F}_{\alpha} & = \left({1/\sqrt{\mu} \atop 0}\ {0 \atop \sqrt{\mu}}\right) \left({\cos \alpha \atop -(1/\mu) \sin \alpha}\ {\mu \sin \alpha \atop \cos \alpha}\right) \left({\sqrt{\mu} \atop 0}\ {0 \atop 1/\sqrt{\mu}}\right)\\
                            & = {\cal M}_{1/\sqrt{\mu}} {\bf M}_{\rm focus} {\cal M}_{\sqrt{\mu}}.
\end{align}
In this case, the effective angle of the FrFT performed by the optical system is $\cos \alpha = 1 - d/f$ \cite{Moreno05}, if we put position in units of $\sqrt{\lambda f}$ (compare section~\ref{sec:FourierLenses}).

When a source is paraxial with the lens ($\cos \theta = 1$) and is located a distance ${\cal D}$ in front of it, its image is found a distance
\begin{equation}
d = f/(1 - f/{\cal D}) \approx f (1 + f/{\cal D})
\end{equation}
behind the lens.  The imaging process is then equivalent to a FrFT with order given by $\cos \alpha \approx -f/{\cal D}$.  We can compare this with the FrFT order given by equation~\ref{eqn:alphaFocus}.  In that case, if the unit $s$ is chosen to be $\sqrt{\lambda f}$, then $\cos \alpha = -\sin \beta \approx -f/{\cal D}$, an identical result.

An interferometer that performs the FrFT on an electric field \emph{focuses} an image.  Conventional interferometers are focused at infinity, as suited for most astronomical sources.  But there is no reason why they cannot focus on things that are closer.

\subsection{Fractional dirty beams, fractional convolutions, and related horrors}
As with regular interferometers (Section~\ref{sec:InterferometryIntro}), measurements become much more complicated with actual interferometers.  First, the measured signal at a given antenna depends on its primary beam's sensitivity in the direction of the source.  Frequently, the primary beams can be approximated as the same for each antenna.  But for a nearby source, the parallax effect implies that the sensitivity of some antennas may be greater than others, if only because the antennas are not all pointed directly at the source.  This could complicate the analysis if the parallax angle is comparable to or larger than the primary beams.  For distant objects with small parallax angles detected by dipole antennas, this may not be as much of an issue.  As in section~\ref{sec:InterferometerComplications}, I will treat the antenna response as a constant $\sqrt{A}$, although this needs to be treated for more realistic situations.

But a fundamental difficulty is the ``filtering'' $S({\bf x})$ performed by the interferometer as it samples the electric signal.  The measured electric field is multiplied by $S({\bf x})$, as in equation~\ref{eqn:SFilteredE}, and the Fractional Fourier Transformed field is 
\begin{equation}
E_{\rm dirty} = {\cal F}_{\alpha} [E^{\prime} (x)] = {\cal F}_{\alpha} [\sqrt{A} E(x) \times S(x)]
\end{equation}
The usual convolution theorem of the Fourier transform does \emph{not} apply (after all, the identity operation is ${\cal F}_0$).  Instead, one must replace the convolution operation with a fractional convolution operation that merely reduces to normal convolution when $\alpha = \pi / 2$.  Then we can say that
\begin{equation}
\label{eqn:EDirtyFrac}
E_{\rm dirty} = {\cal F}_{\alpha} [E(x)] \ast_{-\alpha} P_{\alpha}^{\rm dirty} = {\cal F}_{\alpha} [E(x)] \ast_{\pi - \alpha} P_{\alpha}^{\rm dirty},
\end{equation}
where $P_{\alpha}^{\rm dirty} = {\cal F}_{\alpha} [S]$ is the fractional dirty beam of the interferometer \cite{Ozaktas01}.  Even if $S$ is a simple rectangular pulse, the fractional dirty beam does not have a simple analytic expression and it has imaginary components \cite{Almeida94}. 

One example of a definition for fractional convolution is given by \cite{Zayed98}:
\begin{equation}
(f \ast_{\alpha} g)(\chi) = e^{\pi i \chi^2 \cot \alpha} [(f e^{-\pi i \chi^2 \cot \alpha}) \ast (g e^{-\pi i \chi^2 \cot \alpha})],
\end{equation}
which I have altered to fit this paper's conventions regarding factors of $2 \pi$ (see also \cite{Almeida97}).  Then, one can write
\begin{multline}
E_{\rm dirty} e^{\pi i \chi^2 \cot \alpha} = (\sqrt{A} e^{\pi i \chi^2 \cot \alpha} {\cal F}_{\alpha} [E(\chi)]) \\
                                           \ast (e^{\pi i \chi^2 \cot \alpha} P_{\alpha}^{\rm dirty}),
\end{multline}
which is a standard deconvolution problem.

\subsection{The limits of fractional interferometers \label{sec:FracInterferometerLimits}}
Unless one works at very low frequencies (e.g., studying atmospheric phenomena at kHz frequencies), we are typically in the regime where ${\cal D} \gg \lambda$.  This condition motivates my use of the Taylor series for ${\cal D}^{\prime}/{\cal D}$ in equation~\ref{eqn:PhiTaylor}.  The reader may notice that the order $Q$ of the FrFT is very close to 1 in this case (equation~\ref{eqn:alphaFocus}).  Although one might worry whether the interferometer can measure the phase of the electric field well enough for the FrFT to be needed, it turns out they can for sources within the atmosphere and into cislunar space.

An interferometer can measure the parallax effect of a nearby source if the parallax angle between one end of the array and the other is bigger than the limiting resolution from Rayleigh's criterion.  Another way to think about it is that the frequency derivative of the chirp can be measured if the phase slides more than 1 radian across the array from a constant tone.  This gives us:
\begin{equation}
p \approx \tan p = \frac{L \cos \theta}{{\cal D}} \ga \frac{\lambda}{L}
\end{equation}
for an array of width $L$.  The relevant condition is
\begin{equation}
{\cal D} \la \frac{L^2 \cos \theta}{\lambda} = 330\ \km\ \cos \theta\ \left(\frac{L}{\km}\right)^2 \left(\frac{\nu}{100\ \MHz}\right)
\end{equation}
using values that could be typical of a 21 cm survey.  When using the longest baselines of LOFAR or SKA ($\sim 1000\ \km$), or the highest frequencies of VLA ($\sim 50\ \GHz$), the Moon itself is just out of focus with a blur of $\sim 0.6$ arcsec (about 1 km).  These limits for various arrays are listed in Table~\ref{table:FracInterferometryLimits}, under ${\cal D}_{\rm linear}$.  

\begin{table*}
\begin{tabular}{llllccc}
Array & $L$ & $\lambda$ & $\nu$ & ${\cal D}_{\rm linear}$ & ${\cal D}_{\rm quadratic}$ & References\\
\hline
\multicolumn{7}{c}{Kilometric radio}\\
\hline
Astronomical Low Frequency Array (ALFA) & 100\ \km               & 10\ \km     & 30\ \kHz & 1000\ \km   & 560\ \km                & \cite{Jones00}\\
                                        &                        & 10\ \meter  & 30\ \MHz & $10^6$\ \km & $1.8 \times 10^4$\ \km  & \\
Hobart array                            & 1\ \km                 & 60\ \meter  & 5\ \MHz  & 17\ \km     & 7.2\ \km                & \cite{Ellis66}\\
\hline
\multicolumn{7}{c}{Meter radio}\\
\hline
LOFAR Superterp                         & 240\ \meter            & 10\ \meter  & 30\ \MHz  & 5.8\ \km   & 2.1\ \km                & \cite{vanHaarlem13}\\
                                        &                        & 1.5\ \meter & 200\ \MHz & 9.6\ \km   & 1.9\ \km                & \\
LOFAR Core                              & 3.5\ \km               & 10\ \meter  & 30\ \MHz  & 1200\ \km  & 120\ \km                & \cite{vanHaarlem13}\\
                                        &                        & 1.5\ \meter & 200\ \MHz & 8200\ \km  & 300\ \km                & \\
LOFAR Remote                            & 121\ \km               & 10\ \meter  & 30\ \MHz  & $1.5 \times 10^6\ \km$ & $2.4 \times 10^4\ \km$ & \cite{vanHaarlem13}\\
                                        &                        & 1.5\ \meter & 200\ \MHz & 0.065\ \AU & $6.1 \times 10^4\ \km$  & \\
LOFAR All                               & 1158\ \km              & 10\ \meter  & 30\ \MHz  & 0.90\ \AU  & $7.0\ \times 10^5\ \km$ & \cite{vanHaarlem13}\\
                                        &                        & 1.5\ \meter & 200\ \MHz & 6.0\ \AU   & 0.012\ \AU              & \\
\hline
\multicolumn{7}{c}{Centimeter radio}\\
\hline
GBT                                     & 110\ \meter            & 300\ \cm    & 100\ \MHz  & 4.1\ \km    & 1.2\ \km               & \cite{Prestage09}\\
                                        &                        & 0.26\ \cm   & 115\ \GHz  & 4700\ \km   & 40\ \km                & \\
VLA (D configuration)                   & 1\ \km                 & 30\ \cm     & 1\ \GHz    & 3300\ \km   & 100\ \km               & \cite{Perley11}\\
                                        &                        & 0.6\ \cm    & 50\ \GHz   & $1.7 \times 10^5\ \km$ & 720\ \km    & \\
VLA (A configuration)                   & 36\ \km                & 30\ \cm     & 1\ \GHz    & 0.029\ \AU  & $2.2 \times 10^4\ \km$ & \cite{Perley11}\\
                                        &                        & 0.6\ \cm    & 50\ \GHz   & 1.4\ \AU    & $1.6 \times 10^5\ \km$ & \\
VLBI                                    & $10^4$\ \km            & 20\ \cm     & 1.5\ \GHz  & 0.016\ \pc  & 0.84\ \AU              & \\
                                        &                        & 1\ \cm      & 30\ \GHz   & 0.32\ \pc   & 3.7\ \AU               & \\
RadioAstron + VLBI                      & $3.5 \times 10^5$\ \km & 92.5\ \cm   & 324\ \MHz  & 4.3\ \pc    & 81\ \AU                & \cite{Andreyanov14}\\
                                        &                        & 1.35\ \cm   & 22.2\ \GHz & 290\ \pc    & 670\ \AU               & \\
Interplanetary radio interferometer     & 1\ \AU                 & 20\ \cm     & 1.5\ \GHz  & 3.6\ \Mpc   & 7.4\ \pc               & \\
\hline
\multicolumn{7}{c}{Submillimeter}\\
\hline
Event Horizon Telescope                 & $10^4$\ \km            & 1\ \mm      & 300\ \GHz  & 3.2\ \pc    & 12\ \AU                & \cite{Lu14}\\
\hline
\multicolumn{7}{c}{Near infrared and optical}\\
\hline
Thirty meter telescope                  & 30\ \meter             & 0.6\ $\um$  & 500\ \THz  & 0.010\ \AU  & 380\ \km               & \\
CHARA                                   & 331\ \meter            & 0.6\ $\um$  & 500\ \THz  & 1.2\ \AU    & $1.4 \times 10^4\ \km$ & \cite{tenBrummelaar05}\\
NIR VLBI                                & $10^4$\ \km            & 1\ $\um$    & 300\ \THz  & 3.2\ \kpc   & 370\ \AU               & \\
\hline
\multicolumn{7}{c}{X-rays}\\
\hline
Maxim Pathfinder                        & 1.4\ \meter            & 200\ \pm    & 150\ \PHz  & 6.6\ \AU    & 2100\ \km             & \cite{Cash05}\\
Maxim                                   & 300\ \meter            & 200\ \pm    & 150\ \PHz  & 1.5\ \pc    & 0.044\ \AU            & \cite{Cash05}\\
\end{tabular}
\caption{Closest distances for which normal and fractional interferometry apply for various arrays.  These correspond to the domains where linear ($\la {\cal D}_{\rm linear}$) and quadratic ($\la {\cal D}_{\rm quadratic}$) approximations to the phase with distance (equation~\ref{eqn:PhiTaylor}) are valid.}
\label{table:FracInterferometryLimits}
\end{table*}

But if the source is too close to the array, then the approximation in equation~\ref{eqn:PhiTaylor} is insufficient.   The next term in $\phi$ is $(2 \pi {\cal D}/\lambda) \times [-(1/2) (x/{\cal D})^3 \sin \theta \cos^2 \theta]$.  The quadratic approximation is sufficient only if this next term adds or subtracts less than 1 radian to the phase:
\begin{equation}
\left|\frac{\pi}{{\cal D}} \left(\frac{x}{{\cal D}}\right)^3 \sin \theta \cos^2 \theta\right| \ga 1.  
\end{equation}
We have
\begin{align}
\nonumber {\cal D} & \la L |\cos \theta| \sqrt{\frac{\pi L |\sin \theta|}{\lambda}} \\
\label{eqn:FracInterferometryLimit}
            & = 32\ \km\ |\cos \theta|\ \sqrt{|\sin \theta|}\ \left(\frac{L}{\km}\right)^{3/2} \left(\frac{\nu}{100\ \MHz}\right)^{-1/2}.
\end{align}
The above condition is listed for various arrays in Table~\ref{table:FracInterferometryLimits} too, under ${\cal D}_{\rm quadratic}$.  Even higher $m$th order terms of the approximation are proportional to $x^m / (\lambda {\cal D}^{m-1})$, and are valid for ${\cal D} \la L^{m/(m-1)} \lambda^{-1/(m-1)}$.  The Taylor series breaks down entirely when ${\cal D} \la L$.

\subsection{Subarrays or fractional interferometry?}
Fractional interferometry corresponds to a second-order approximation of the phase with position, or a first-order approximation in spatial frequency.  This seems obviously advantageous over normal interferometry, which is a first-order approximation of the phase (zeroth-order in frequency) -- we fit the signal with a line of arbitrary slope instead of a zero-slope line in Wigner space (Figure~\ref{fig:WignerFrFT}).  But another way to describe a signal is to break the array into many smaller arrays and fit the signal in Wigner space with a series of constant values, one for each sub-array.  Likewise, one could try to fit a signal with a nonzero second derivative of $\kappa$ by breaking the signal into a series of lines, to get around the limits imposed by equation~\ref{eqn:FracInterferometryLimit}.

Essentially, the problem is to minimize the angular beam width $W$ if the source may be out of focus.  Consider a 1D subarray centered at $x = x_0$ with a total width of $l$.  There are two terms in $W$.  The first is the resolution limit imposed by diffraction, $W_{\rm diff} = \lambda / l$, which favors large $\ell$.  The second is the apparent angular size of the source's image, $W_{\rm image}$, which gets larger as the source goes out of focus.  $W_{\rm image}$ is basically the projected ``width'' (or overhang) of the signal in Wigner space after the interferometer performs a FFT or FrFT; it is calculated as the overhang in $x^{\prime}$ (which is almost, but not technically, $\kappa$) in units of $1/\lambda$.  With a standard interferometer, the best-fit linear approximation to the phase leaves a residual signal width of
\begin{equation}
W_{\rm image}^{\rm linear} \approx \frac{l \cos^2 \theta}{2 {\cal D}} \left(3 \frac{x_0}{{\cal D}} \sin \theta + 1\right) \approx \frac{l \cos^2 \theta}{2 {\cal D}}
\end{equation}
for a distant source.  (Note that $x_0$ ranges from $-l/2$ to $l/2$.)  Minimizing this term favors small $l$.  Then, supposing the total beam width is about $W \approx \min[W_{\rm diff}, W_{\rm image}^2]$, the optimal beam width is,
\begin{equation}
W_{\rm best}^{\rm linear} \approx \cos \theta \sqrt{\frac{\lambda}{2 {\cal D}}}
\end{equation}
for an array size of
\begin{equation}
l_{\rm best}^{\rm linear} \approx \sec \theta \sqrt{2 \lambda {\cal D}}.
\end{equation}
Using a larger subarray (or telescope) than $l_{\rm best}^{\rm linear}$ actually makes the image \emph{blurrier} because the parallax of the source across the array is detectable.

The best-fit quadratic approximation to the phase leaves a residual image width of
\begin{equation}
W_{\rm image} = \frac{3}{8} \left(\frac{l}{\cal D}\right)^2 \sin \theta \cos^2 \theta.
\end{equation}
This allows an optimal width of 
\begin{equation}
W_{\rm best}^{\rm quadratic} \approx \left(\frac{3}{8} \frac{\lambda^2}{{\cal D}^2} \sin \theta \cos^2 \theta\right)^{1/3}
\end{equation}
for an array size of
\begin{equation}
l_{\rm best}^{\rm quadratic} \approx \left(\frac{8}{3} \lambda {\cal D}^2 \csc \theta \sec^2 \theta \right)^{1/3}.
\end{equation}
Again, a larger subarray than $l_{\rm best}^{\rm quadratic}$ just makes things worse.

Using a FrFT instead of a Fourier transform improves the image angular width as
\begin{equation}
\frac{W_{\rm best}^{\rm quadratic}}{W_{\rm best}^{\rm linear}} \approx \left(\frac{9}{8} \frac{\lambda}{\cal D} \tan^2 \theta \right)^{1/6},
\end{equation}
for subarrays with the optimal size, where
\begin{equation}
\frac{l_{\rm best}^{\rm quadratic}}{l_{\rm best}^{\rm linear}} \approx \left(\frac{8}{9} \frac{\cal D}{\lambda} \cot^2 \theta \right)^{1/6}.
\end{equation}
Although these ratios evolve slowly with distance, note that the relevant factor of ${\cal D}/\lambda$ is very big.  As a result, the fractional interferometer can improve the best possible beam width by orders of magnitude.

\begin{table*}
\begin{tabular}{lllcccccc}
Source & ${\cal D}$ & $\lambda$ & \multicolumn{3}{c}{Standard interferometer} & \multicolumn{3}{c}{Fractional interferometer}\\
       &            &           & $l_{\rm best}^{\rm linear}$ & $W_{\rm best}^{\rm linear}$ & $\Delta S_{\rm best}^{\rm linear}$ & $l_{\rm best}^{\rm quadratic}$ & $W_{\rm best}^{\rm quadratic}$ & $\Delta S_{\rm best}^{\rm quadratic}$\\
\hline
Thunderstorm             & 10\ \km                 & 10\ \meter & 0.4\ \km             & 1$^{\circ}$ & 0.2\ \km             & 1\ \km               & 0.4$^{\circ}$ & 70\ \meter \\
ISS                      & 500\ \km                & 30\ \cm    & 0.5\ \km             & 2\ \arcmin  & 0.3\ \km             & 6\ \km               & 10\ \arcsec   & 30\ \meter\\
Geosynchronous satellite & $3.6 \times 10^4\ \km$  & 1\ $\um$   & 8\ \meter            & 20\ \mas    & 4\ \meter            & 2\ \km               & 0.1\ \mas     & 2\ \cm\\
Moon                     & $3.84 \times 10^5\ \km$ & 1\ \meter  & 30\ \km              & 7\ \arcsec  & 10\ \km              & 700\ \km             & 0.3\ \arcsec  & 0.5\ \km\\
                         &                         & 1\ $\um$   & 30\ \meter           & 7\ \mas     & 10\ \meter           & 7\ \km               & 30\ \uas      & 5\ \cm\\
Neptune                  & 30\ \AU                 & 1\ \meter  & 3000\ \km            & 70\ \mas    & 2000\ \km            & $4 \times 10^5\ \km$ & 0.5\ \mas     & 10\ \km\\
                         &                         & 1\ $\um$   & 3\ \km               & 70\ \uas    & 2\ \km               & 4000\ \km            & 60\ \nas      & 1\ \meter\\
Solar neighborhood       & 10\ \pc                 & 1\ \meter  & $8 \times 10^5\ \km$ & 0.3\ \mas   & $4 \times 10^5\ \km$ & 4\ \AU               & 0.3\ \uas     & 500\ \km\\
                         &                         & 1\ $\um$   & 800\ \km             & 0.3\ \uas   & 400\ \km             & 0.04\ \AU            & 30\ \pas      & 50\ \meter\\
Galactic Center          & 8\ \kpc                 & 1\ \meter  & 0.1\ \AU             & 9\ \uas     & 0.07\ \AU            & 400\ \AU             & 4\ \nas       & 5000\ \km\\
                         &                         & 1\ \mm     & $7 \times 10^5\ \km$ & 0.3\ \uas   & $4 \times 10^5\ \km$ & 40\ \AU              & 40\ \pas      & 50\ \km\\
                         &                         & 1\ $\um$   & $2 \times 10^4\ \km$ & 9\ \nas     & $1 \times 10^4\ \km$ & 4\ \AU               & 0.4\ \pas     & 0.5\ \km\\
                         &                         & 20\ \nm    & 3000\ \km            & 1\ \nas     & 2000\ \km            & 1\ \AU               & 30\ \fas      & 30\ \meter\\
M87                      & 20\ \Mpc                & 1\ \meter  & 7\ \AU               & 0.2\ \uas   & 4\ \AU               & 0.3\ \pc             & 20\ \pas      & $6 \times 10^4\ \km$\\
                         &                         & 1\ \mm     & 0.2\ \AU             & 6\ \nas     & 0.1\ \AU             & 0.03\ \pc            & 0.2\ \pas     & 600\ \km\\
                         &                         & 1\ $\um$   & $1 \times 10^6\ \km$ & 0.2\ \nas   & $5 \times 10^5\ \km$ & 700\ \AU             & 2\ \fas       & 6\ \km\\
                         &                         & 20\ \nm    & $2 \times 10^5\ \km$ & 30\ \pas    & $8 \times 10^4\ \km$ & 200\ \AU             & 0.2\ \fas     & 0.5\ \km\\
Distant galaxies         & 2\ \Gpc                 & 1\ \meter  & 70\ \AU              & 20\ \nas    & 40\ \AU              & 7\ \pc               & 1\ \pas       & $3 \times 10^5\ \km$\\
                         &                         & 1\ \mm     & 2\ \AU               & 0.6\ \nas   & 1\ \AU               & 0.7\ \pc             & 10\ \fas      & 3000\ \km\\
                         &                         & 1\ $\um$   & 0.07\ \AU            & 20\ \pas    & 0.04\ \AU            & 0.07\ \pc            & 0.1\ \fas     & 30\ \km\\
\end{tabular}
\caption{Best possible angular resolutions with a standard interferometer (first order in phase) and a fractional interferometer (second order in phase).  The largest resolvable features on the source is denoted by $\Delta S$.  All trigonometric terms on $\theta$ are assumed to be $\sim 1$.}
\label{table:OptimalResolutions}
\end{table*}

I demonstrate this by calculating the best possible image resolutions with standard and fractional interferometry for various sources in Table~\ref{table:OptimalResolutions}.  For sources within the Solar System, standard interferometers have blurred, sub-optimal images for VLBI baselines.  In addition, if a RadioAstron-like baseline is ever added to the EHT, standard interferometry cannot resolve scales below $\sim 1\%$ of the Galactic Center's supermassive black hole Schwarzschild radius.  The fractional Fourier interferometer performs much better.  One would need baselines that are spread across the Solar System before one would need to go to third order, if the sources are interstellar.  Fractional interferometers are sufficient for our present astronomical needs.

\subsection{An example calculation of a fractional interferometer}
I shall now demonstrate how much using a FrFT instead of a simple Fourier transform matters for the beam size.  Consider a 1D FFTT that is $L = 1\ \km$ wide designed to observe GHz radio waves.  It has $2^{15} = 32768$ elements, spaced $3.05\ \cm$ apart.  It therefore has enough information to sample $1\ \GHz$ radiation ($\lambda = 29.98\ \cm$).  I consider a case without any noise for simplicity.

Now let's say that one wishes to focus on a $1\ \GHz$ emitter that is ${\cal D} = 500\ \km$ away at a zenith angle of $45^{\circ}$.  The transmitter may be a low earth orbit satellite, like the International Space Station (ISS), but it is a point source.  According to equation~\ref{eqn:alphaFocus}, the optimal FrFT has order very near 1, with $\Delta Q = Q - 1 = 1.91 \times 10^{-7}$.  How much of a difference does this make?

\begin{figure*}
\includegraphics[width=8cm]{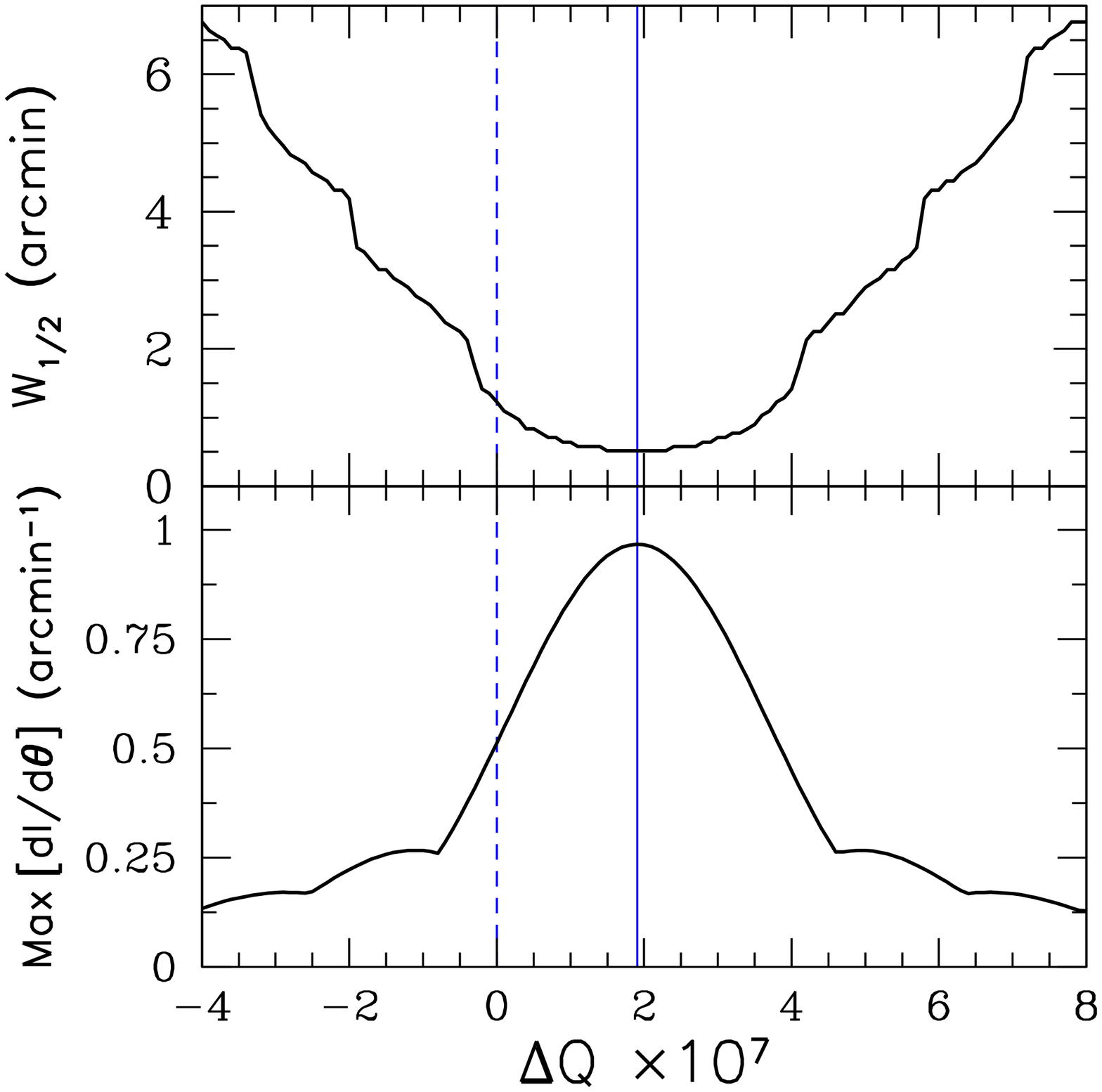}\,\includegraphics[width=8cm]{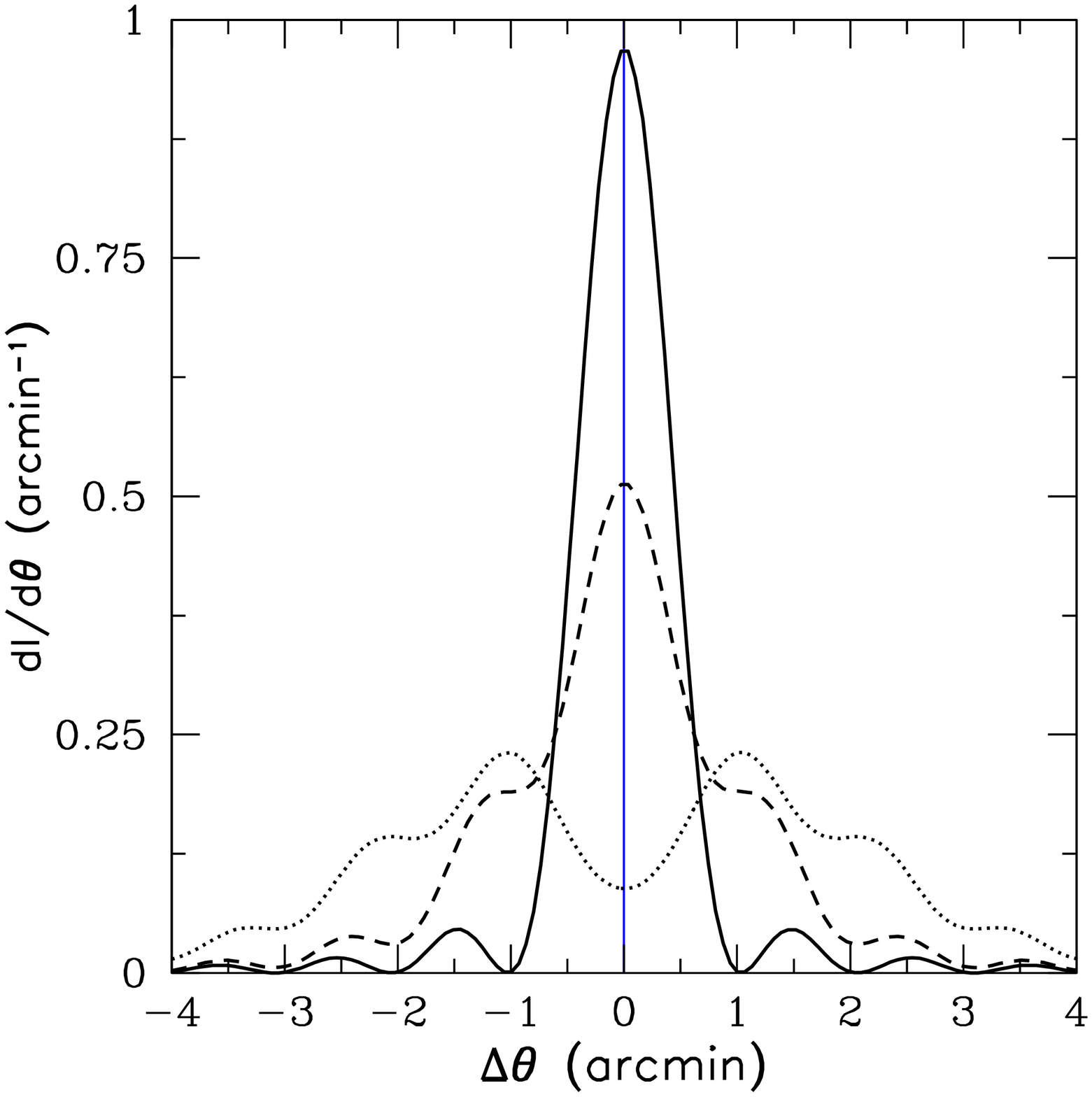}
\caption{How the characteristics of the image of the transmitter vary with FrFT order $\Delta Q$ (left), including the half-power beam width and the maximum intensity.  On right, the intensity distributions of the image; plotted images are for $\Delta Q = 1.9 \times 10^{-7}$ (solid), $\Delta Q = 0$ (dashed), and $\Delta Q = -1.9 \times 10^{-7}$ (dotted).  The signal is normalized so that the total detected power is 1.\label{fig:ISSTransientImages}}
\end{figure*}

We can measure the effective image size by considering the smallest range of $x^{\prime}$ around the point of maximum intensity that contains 50\% of the power.  This quantity, dubbed $W_{1/2}$, is plotted in Figure~\ref{fig:ISSTransientImages}.  There is a flat bottom around the optimal $\Delta Q$, but the image width begins increasing quickly as the deviation from this best $\Delta Q$ passes $(0.5$--$1) \times 10^{-7}$.  A simple Fourier transform double the image size, and cuts the peak intensity in half.  Note how the intensity peak is smeared out by the normal Fourier transform compared with the FrFT in the right panel.

The fractional interferometer clearly performs better than the normal FFTT analysis, doubling the signal even for this tiny of a $\Delta Q$.

\subsection{Targets for a fractional interferometer}

A fractional interferometer can increase an interferometer's sensitivity and ability to maps sources that are relatively nearby.  The inability of conventional interferometers to ``focus'' on nearby sources has been understood for a long time.  Various fixes for focusing on these objects have been proposed, including adding a quadratic time delay for different elements (\cite{Fukushima94Sovers98}; see also \cite{Carter89}), and using a nonlinear mapping on the $uv$ plane \cite{Duev12}.  The FrFT provides a natural framework for solving the problem.

As seen in Table~\ref{table:FracInterferometryLimits}, the entire Solar System is out of focus in radio VLBI arrays.  Yet VLBI is an important tool for tracking space probes within the Solar System.  Aside from simply keeping tabs on where the probes are, precise measurements of space probes are useful for pinning down the locations of the planets \cite{Jones11}, performing tests of general relativity and modified gravity \cite{Hees14}, and monitoring the winds of planets and moons with atmospheric probes \cite{Sagdeyev92,Witasse06}.  Among the objects studied with VLBI tracking are Venus \cite{Sagdeyev92}, Mars \cite{Hildebrand94}, Saturn \cite{Jones11}, and Titan \cite{Witasse06}.  

The Moon is also a frequent target of VLBI observations.  The later Apollo missions planted radio beacons on the Moon's surface as part of the Apollo Lunar Surface Experiments Packages (ALSEPs).  Early selenometry benefited from VLBI observations of these beacons, which pinned down their relative positions and constrained the Moon's motion \cite{Counselman73King76}.\footnote{The ALSEPs were turned off in the 1970s, but there's no reason why future crewed or uncrewed missions to the Moon could not place additional radio transmitters.}  VLBI tracking of lunar satellites like SELENE allowed for the determination of the Moon's gravitational field to high order \cite{Goossens11}.  

Artificial satellites in geosynchronous and low earth orbit are out of focus in normal VLBI observations, as well as in high-frequency observations by the Very Large Array (Table~\ref{table:FracInterferometryLimits}).  These artificial satellites are monitored on occasion with long baseline radio interferometry \cite{Preston72}.  There has also been recent interest in using optical interferometry to directly observe communications satellites in geosynchronous orbit, to diagnose their status \cite{Armstrong09Hindsley11}.  Yet, according to Table~\ref{table:FracInterferometryLimits}, optical observations of these satellites are out of focus, and FrFT techniques must be applied to get a sharper image.  

Even closer to home, there are many radio phenomena in the atmosphere that could benefit from fractional interferometry.  Meteors vaporize into plasma in the atmosphere, which reflects artificial radio signals \cite{Lazio10}.  Meteors can emit radio emission on their own, too \cite{Obenberger14}.  Lightning frequently is a source of radio emission from Extremely Low Frequencies ($\la\ \kHz$) up through hundreds of MHz \cite{LeVine80Holden95,Rodger06}.  Sprites, which accompany some thunderstorms, also are sources of Very Low Frequency emission \cite{Rodger06}.  These kinds of electrical phenomena excite sferics and whistlers in the ionosphere at frequencies below a MHz \cite{Helliwell65}.  Finally, Ultra High Energy Cosmic Rays (UHECRs) produce particle showers when they interact with the atmosphere.  The detection of radio pulses accompanying these showers is an important technique for measuring the UHECR flux at the highest energies \cite{Lehtinen04,Gorham09}.  The high frequency radio transients in the atmosphere are of interest to the numerous radio astronomy arrays, and interferometry is occasionally used to study the low frequency transients associated with lightning \cite{Rhodes94Mezentsev13}.  Fractional interferometers of the right frequency range may be more sensitive to all of these phenomena and could potentially map the shapes of the radio-emitting regions, perhaps providing new diagnostics.  

Observations of the new class(es) of Fast Radio Bursts (FRBs) are a good example of what a fractional interferometer might accomplish.  These millisecond-long pulses observed at GHz frequencies are dispersed in a way consistent with being located outside of the Milky Way \cite{FRBs}.  But the situation is confused by the recent discovery of perytons, atmospheric radio transients of similar dispersion, frequency, and duration \cite{BurkeSpolaor11}.  If a FFTT could catch some of these pulses, it could discern whether the FRBs are really astronomical or if they are atmospheric easily.  An atmospheric FRB is out of focus \cite{Kulkarni14}, and its image would be sharpened by a FrFT.  Extraterrestrial FRBs, on the other hand, are essentially at infinity.  

Perhaps the most prosaic, but useful, application of the fractional interferometry technique would be to remove radio frequency interference (RFI) from human activities \cite{Fridman01,Offringa13}.  RFI is generated by electronic devices, both on the ground and in orbit.  RFI is an ever-increasing threat to radio astronomy, and must be removed from radio data \cite{Fridman01}.  In many cases listed in Table~\ref{table:FracInterferometryLimits}, namely those with long baselines or at high frequency, the RFI sources are close enough to be out of focus.  Focusing an interferometer could produce a sharper image of the RFI source, allowing it be removed more cleanly from the data without losing the wanted signal.

\section{The uses of evanescent waves \label{sec:Evanescence}}
In the last section, I assumed that the radiation is high frequency, with both the array and the distance to the source much bigger than the wavelength.  But at very low frequencies, it is quite possible that the array and the source distance are smaller than one wavelength.  This could happen when observing at kHz frequencies, where the wavelength is $30\ \km\ (\nu / 10\ \kHz)^{-1}$, a common regime for sferic studies.  Then we need to consider a near-field limit of microscopy.  The recent advances in superlenses are applicable to this regime.

\subsection{Digital superlenses: perfect radio microscopes}

As I noted in Section~\ref{sec:MatrixOptics}, digital optics are not bound by normal physical constraints.  Digital telescopes can simulate elements with negative index of refraction as easily as elements with positive index of refraction.  Negative index of refraction materials (NIMs) are especially interesting from an optical point of view, as they can be used to construct superlenses.

Superlenses are lenses that can image objects that are smaller than the wavelength of the light they scatter as long as they are in the near-field limit (less than one wavelength away) \cite{Pendry00}.  Even a simple slab of NIM can act as a superlens \cite{Veselago68}.  When light enters a NIM, it bends \emph{away} from the surface normal, according to Snell's Law.  This allows them to focus light both inside the material and again after it leaves the NIM slab \cite{Pendry00}.

The overall effect of a NIM is to effectively undo the propagation of light \cite{Smith00}.  The reverse propagation applies not only to the typical oscillating waves of light, but also exponentially decaying evanescent waves.  The evanescent waves contain information about their source on scales below the wavelength.  They vanish in the far-field ($d \ga \lambda$), but if they enter a NIM before that, they are amplified back up to their original strength or beyond.  Thus, it is possible to reconstruct the electric field at the source in precise detail with a NIM.  This superlens effect allows an optical system to capture information below the diffraction limit in the near-field limit \cite{Pendry00,Zhang08}.

Physical microwave superlenses have been built out of metamaterials, but there are fundamental limits to their performance that do not apply to digital optics.  The most important is causality.  A wavefront cannot bend away from the surface normal instantaneously; the far side of the wavefront needs time to catch up to the near side.  Light in a NIM has a superluminal phase velocity, but the group velocity is necessarily less than $c$, and there is a delay time before the wavefronts are launched \cite{Foteinopoulou03}.  The constraint of causality necessarily introduces dispersion, however, so a physical superlens is limited to a narrow wavelength range \cite{Smith00}.

But a digital system is not limited by causality.  When a digital optical system is synthesized and applied to stored data, the entire history of the electric field is already known, so an acausal system is possible.  Causality is respected in a fundamental sense, because it already takes a light-crossing time for the information to be received by a central processor.  As such, a digital superlens may be designed to have no dispersion and can work at any frequency.  However, the resolution of a digital superlens is still limited by the sampling of the electric field \cite{Smith03}.

Digital superlenses are practical in the near-field regime, when the source is less than one wavelength away.  While this does not apply for astronomical sources, it does apply to atmospheric radio phenomena at very low frequencies.  

Astronomical sources are, of course, so far away that evanescent waves have long since decayed away.  Trying to amplify these vanished waves instead merely amplifies random noise.  The one proposal to use NIMs in astronomical telescopes involves artificially creating evanescent waves in a NIM that simulate the evanescent waves that exist in the near-field of a source, which usually requires some knowledge of what the source looks like \cite{May04}.  An interferometer with a digital superlens is really a \emph{microscope} rather than a telescope.

\subsection{Laplacian interferometers: the (practically) impossible dream}

\begin{figure*}
\includegraphics[width=6cm]{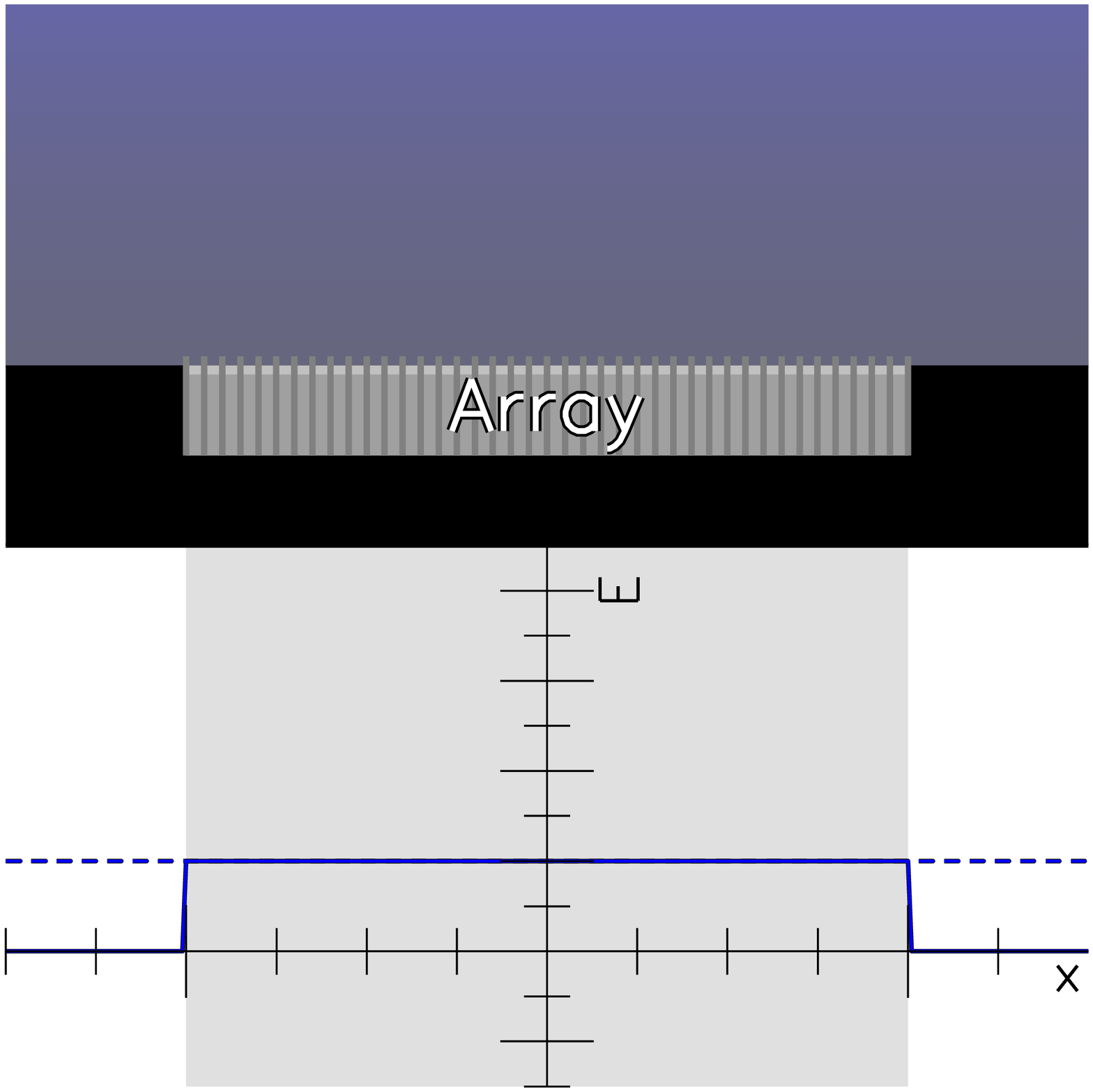}\,\includegraphics[width=6cm]{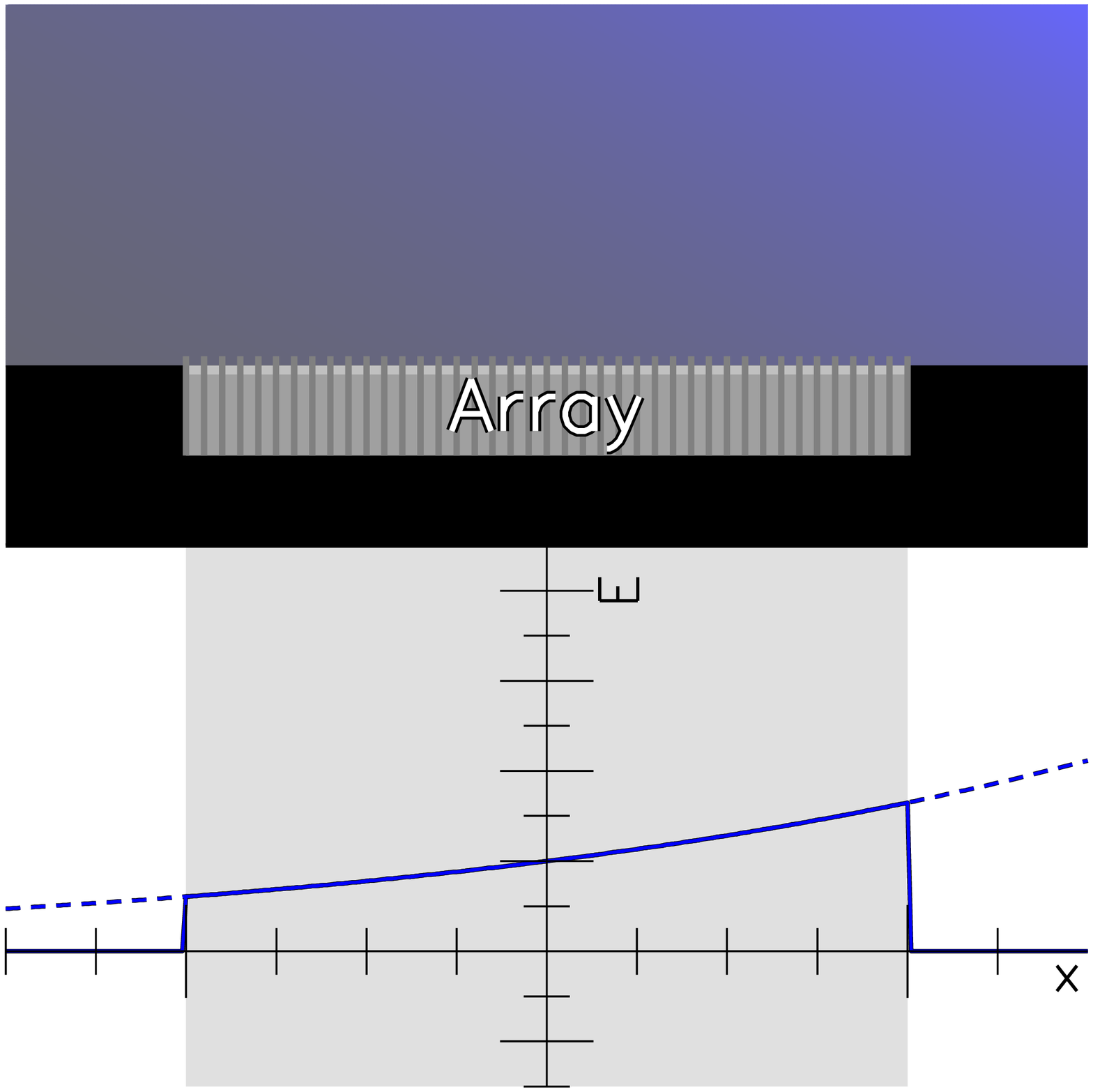}\,\includegraphics[width=6cm]{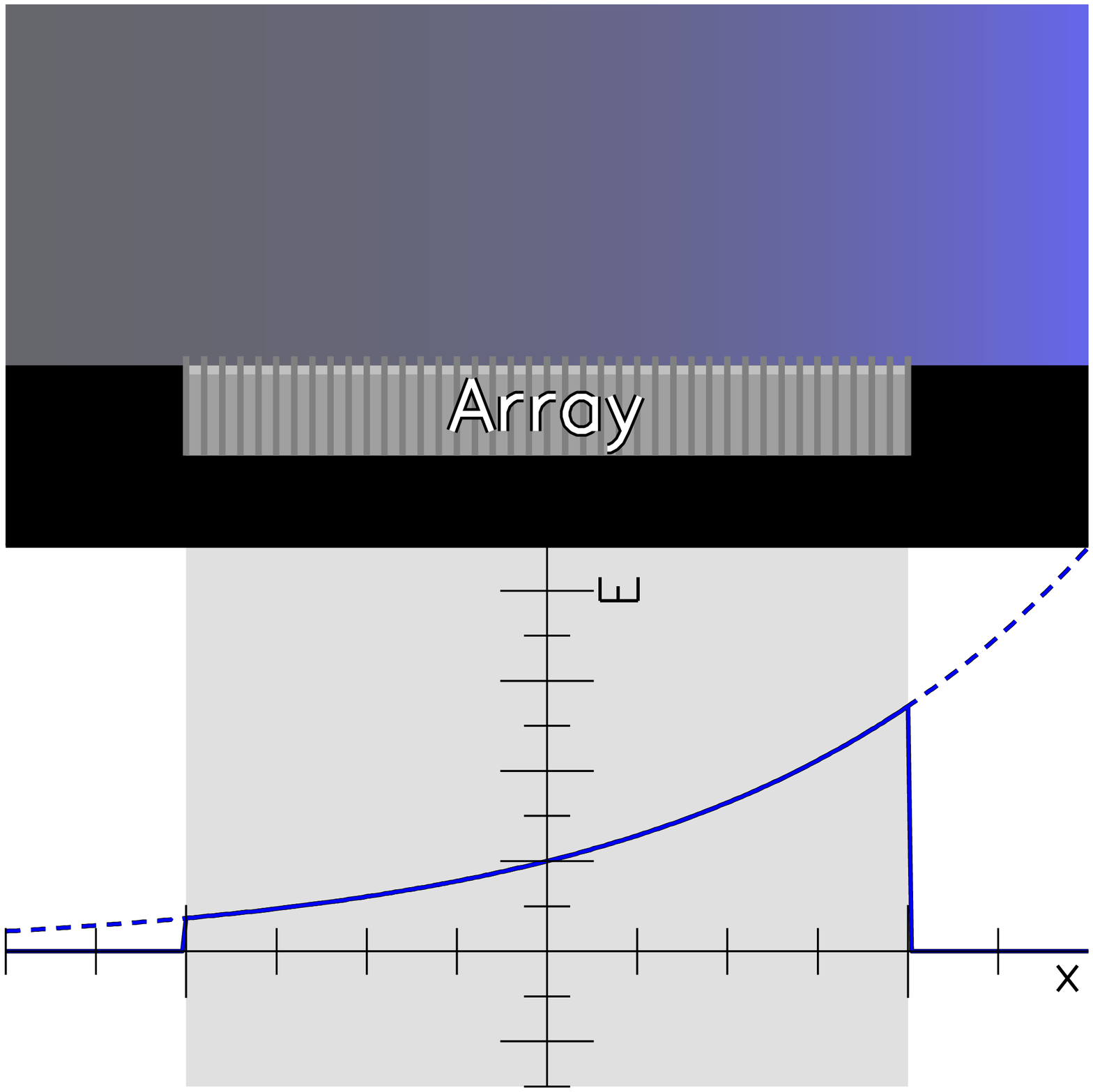}
\caption{Projection effects apply to a static electric field that has exponential scaling towards some direction in the sky.  Separating these kinds of exponentials is impossible in practice.  The lines styles are the same as in Figure~\ref{fig:Interferometry}.\label{fig:LaplacianInterferometry}}
\end{figure*}

If digital superlenses use evanescent waves to reconstruct sub-diffraction images of near-field objects, it is natural to wonder if the evanescent waves can be detected directly.  The evanescent waves are exponentially damped with distance from the source.  The array measures a slice of these fields, which generically has an exponential dependence on position due to projection (Figure~\ref{fig:LaplacianInterferometry}).  This is conceptually similar to normal interferometers, except that the exponentials of imaginary frequency are replaced by exponentials of real frequency.  The necessary transform is the Fourier transform at imaginary frequency -- the inverse Laplace transform.  Can we build a \emph{Laplacian} interferometer around this principle, instead of the usual Fourier interferometer?

There are two types of Laplace transform that we can consider.  The LCT formulation of the Laplace transform, given by
\begin{equation}
{\cal L}_2 = \left({0 \atop i}\ {i \atop 0}\right),
\end{equation}
actually is the bilateral Laplace transform (aside from a phase factor):
\begin{equation}
F(s) = {\cal L}_2 [f(\kappa)] = \sqrt{-i} \int_{-\infty}^{\infty} e^{-2\pi s \kappa} f(\kappa) d\kappa.
\end{equation}  
We also have the LCT version of the inverse bilateral Laplace transform:
\begin{equation}
f(\kappa) = {\cal L}_2^{-1} [F(s)] = \sqrt{-i} \int_{-\infty}^{\infty} e^{2\pi s \kappa} F(\kappa) d\kappa.
\end{equation}  
A single evanescent wave has a single projected scale $\kappa_0$, and has an electric field represented as a spike $\delta(\kappa - \kappa_0)$.  The spike is Laplace transformed into an exponential in $s$-space, $\exp(-2\pi s \kappa_0)$.  Ideally, we would just invert the measured electric field back into a delta function in $\kappa$-space.  Yet the inversion fails because the inverse bilateral Laplace integral diverges.  Bilateral Laplace transforms generally have limited domains of convergence in $\mathbb{C}$.

The one-sided Laplace transform,
\begin{equation}
F(s) = {\cal L} [f(\kappa)] = \int_0^{\infty} e^{-s \kappa} f(\kappa) dk
\end{equation}
has better convergence properties.  Once again, the Laplace transform of $\delta(\kappa - \kappa_0)$ is an exponential in $s$-space, $\exp(-s \kappa_0)$, as long as $\kappa_0 > 0$.  An exponential can be inverse Laplace transformed back into a $\delta$-function.  Conceivably, all we need to do to build an evanescent wave interferometer is perform a one-sided inverse Laplace transform on the measured electric field.  

This is the infamous problem of separating exponentials that crops up from time to time in various fields \cite{Istratov99}.  The underlying obstacle is that that the inversion requires integration in the complex $s$ domain:
\begin{equation}
f(\kappa) = {\cal L}^{-1} [f(\kappa)] = \lim_{C \to \infty} \int_{\gamma - iC}^{\gamma + iC} e^{s \kappa} F(s) ds.
\end{equation}
The contour of integration is the Bromwich contour, for which $\gamma$ has a greater real value than all singularities of $F(s)$.  While this is fine if we know the behavior of $F(s)$ at complex $s$ (as we do for the analytic exponential function), an array can measure the electric field only at real-valued positions.  

Because of the importance of the problem, several numerical algorithms have been developed that attempt to fit exponential(s) to data at real-valued positions \cite{Istratov99}.  For example, one might fit the data with an analytic function (such as Laguerre or Chebyshev polynomials) and then extrapolate it into the complex domain \cite{Lanczos56}.  The exponentials can also be fit by transforming the variables and performing a certain deconvolution \cite{Gardner59}.   

All techniques suffer from extreme instabilities in the presence of noise, because the separation of exponentials is ``ill-posed'' with several solutions \cite{Lanczos56,McWhirter78,Istratov99}.  In functional analysis terms, the exponential functions are far from orthogonal \cite{Lanczos56,Provencher76}.  Since exponentials vary rapidly, even a slight fluctuation requires huge changes in the spectrum in order for the exponentials to cancel out in most places.  Even the number of exponentials that produce the signal can be impossible to extract \cite{Lanczos56}.

The inability to infer which exponentials contribute to the electric field is fundamentally related to the diffraction limit in optics \cite{McWhirter78}, and why we cannot detect evanescent waves from astronomical sources.  An image of an object is formed by an optical system with some entrance pupil.  The image is a Fourier transform of the electric field near the pupil; therefore, the finite size of the pupil filters out high spatial frequency information.  But the Fourier transform necessarily produces an analytic function, so all of the information can hypothetically be reconstructed from the small sliver that passes the pupil \cite{Slepian61}.  The apparently infinite resolution compared to the limited resolution of actual optics presents an apparent paradox \cite{ToraldoDiFrancia69,McWhirter78}.

The paradox is solved by noting that although the high frequency information is still mathematically present, it is suppressed to nearly zero amplitude.  In more formal terms, the eigenvalues of the optical system's eigenfunctions drop from almost exactly one to almost exactly zero as one passes the diffraction bound \cite{ToraldoDiFrancia69,McWhirter78} (see \cite{Piestun00} for a review).  Attempting to find the strengths of the high frequency information results in a division by (nearly) zero, and even the slightest noise results in wild deviations.  The problem is a generic one for integral equations, including the Laplace transform, which smooths the exponential spectrum on small scales \cite{McWhirter78,Provencher82}.  

Physically, the suppression of the high eigenfunctions comes from the decay of evanescent waves decay after propagating any significant length.  In order to measure those waves, they have to be multiplied back up by an exponential factor to their original strength -- or, equivalently, their amplitudes must be divided by nearly zero.  The inevitable noise within the system is also amplified with the evanescent waves, scrambling any information they contain.  Even if we had a perfect detector, quantum fluctuations inject unavoidable noise, limiting the resolution in the far-field \cite{Kolobov00}.  Likewise, the throughput of the optical system is very small for evanescent waves, and any detectable signal requires a vast number of photons with a vast amount of energy \cite{Yu70Zheludev08}.  Or, reversing time, an antenna can broadcast radiation in an arbitrarily narrow angle only at the cost of vast input power \cite{ToraldoDiFrancia52}.

For astronomical objects, with distances\footnote{The Moon at 1 MHz.} of $\gg 10^{5} \lambda$, the evanescent waves are suppressed by a factor $\gg e^{10^5}$.  Hence, we cannot use superlenses to image astronomical objects with arbitrary resolution.   Only in the near-field, where the exponential functions are well-behaved and the evanescent waves have not yet decayed, can the Laplace transform be inverted.  The virtual impossibility of a Laplacian interferometer makes the connection between the two problems explicit.

\section{Beyond the Fourier Basis in Time Domain}

\subsection{Generic waveforms in the time domain \label{sec:TimeDomain}}
Just as non-Fourier transforms can be used for the spatial structure of the electric field, the temporal structure of the field can be decomposed using other transforms.  Of course, there is nothing inherently wrong with using the Fourier transform, as any function can be stably represented as a sum of Fourier modes.  The Fourier transform is not as natural, though, for representing signals whose frequency changes rapidly in time (as noted in \cite{Gorham09,RomeroWolf13}).

Suppose the radiation from a source is isotropic.  We denote the electric field at a distance ${\cal D}^{\prime}$ (see Figure~\ref{fig:FrFTGeometry}) and a time $t$ as $E({\cal D}^{\prime}, t)$.  Because the radiation is isotropic,
\begin{equation}
E({\cal D}^{\prime}, t) = \frac{f(t^{\prime})}{{\cal D}^{\prime}} = \frac{f(t - {\cal D}^{\prime}/c)}{{\cal D}^{\prime}},
\end{equation}
where $f(t^{\prime})$ can be any function with finite energy.  Now we wish to understand how the $f$ waveform varies across the array.  Let $t_0 = t - {\cal D} / c$.  Then,
\begin{equation}
\label{eqn:tGenericInterferometer}
t - \frac{{\cal D}^{\prime}}{c} = t_0 + \frac{({\cal D} - {\cal D}^{\prime})}{c} \approx t_0 - \frac{x}{c}\left[\sin \theta + \left(\frac{x}{{\cal D}}\right) \frac{\cos^2 \theta}{2}\right]
\end{equation}
using the same logic as we used to derive the fractional interferometer in section~\ref{sec:FracInterferometer}.  

\begin{figure*}
\includegraphics[width=6cm]{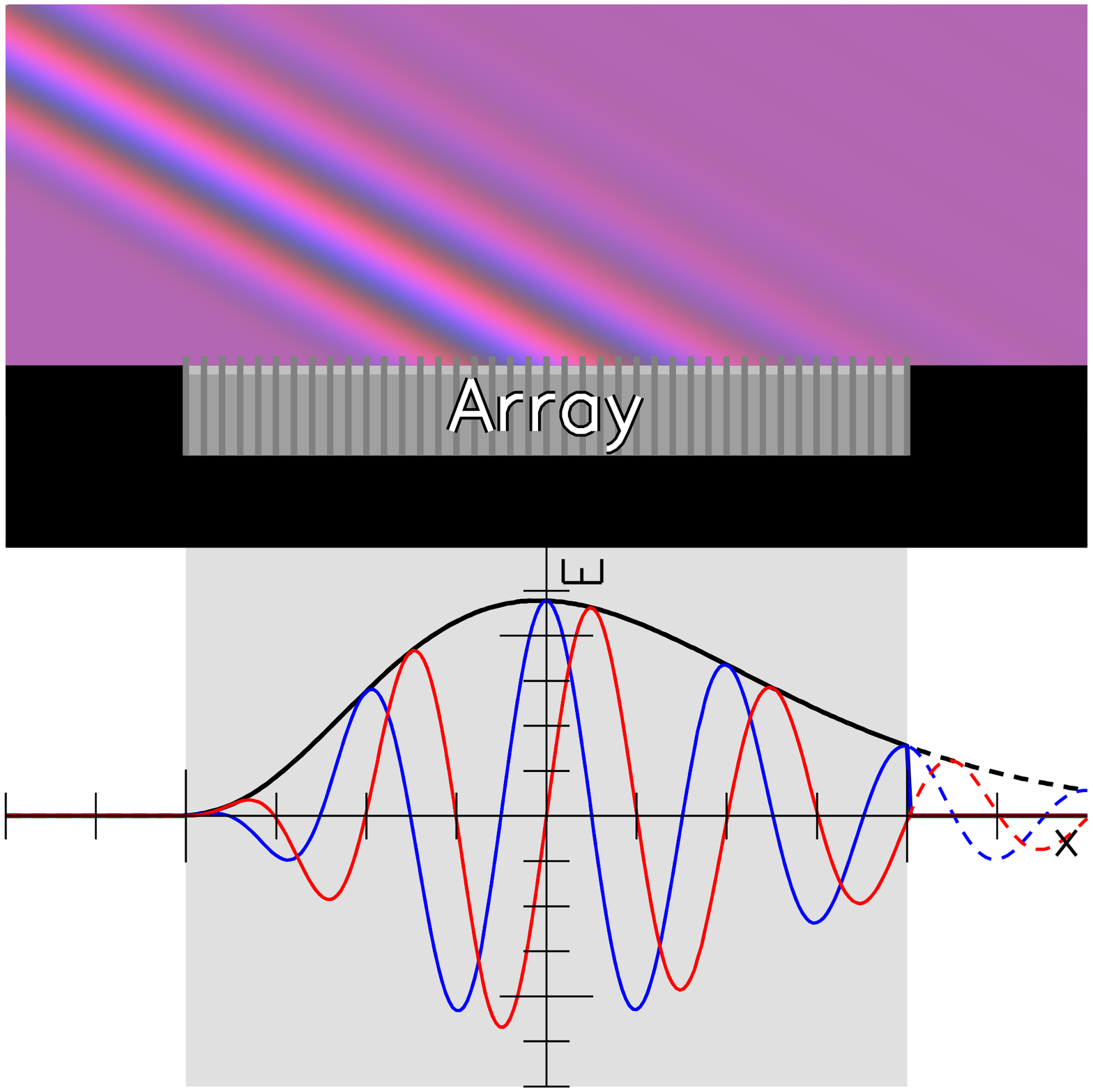}\,\includegraphics[width=6cm]{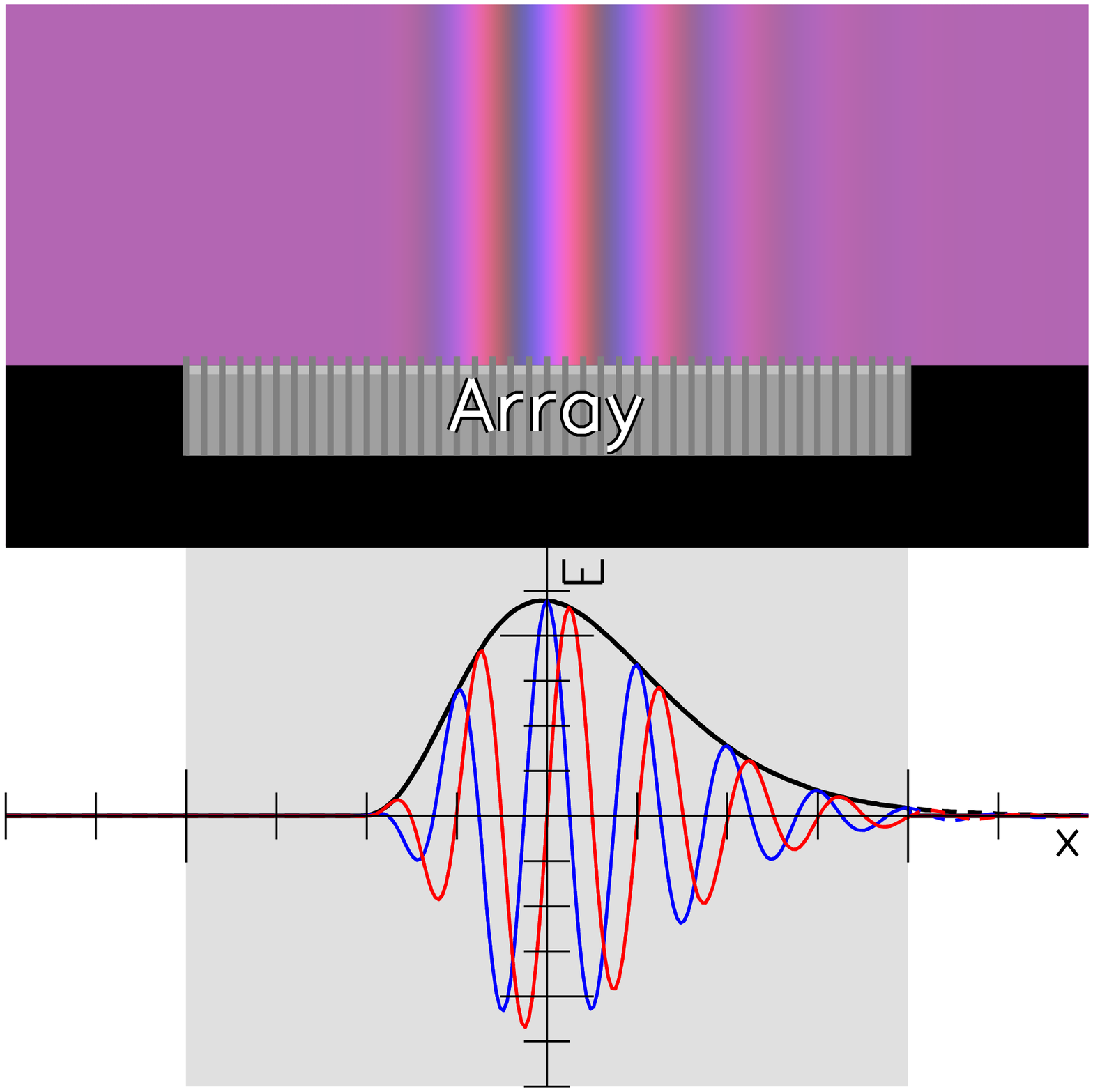}\,\includegraphics[width=6cm]{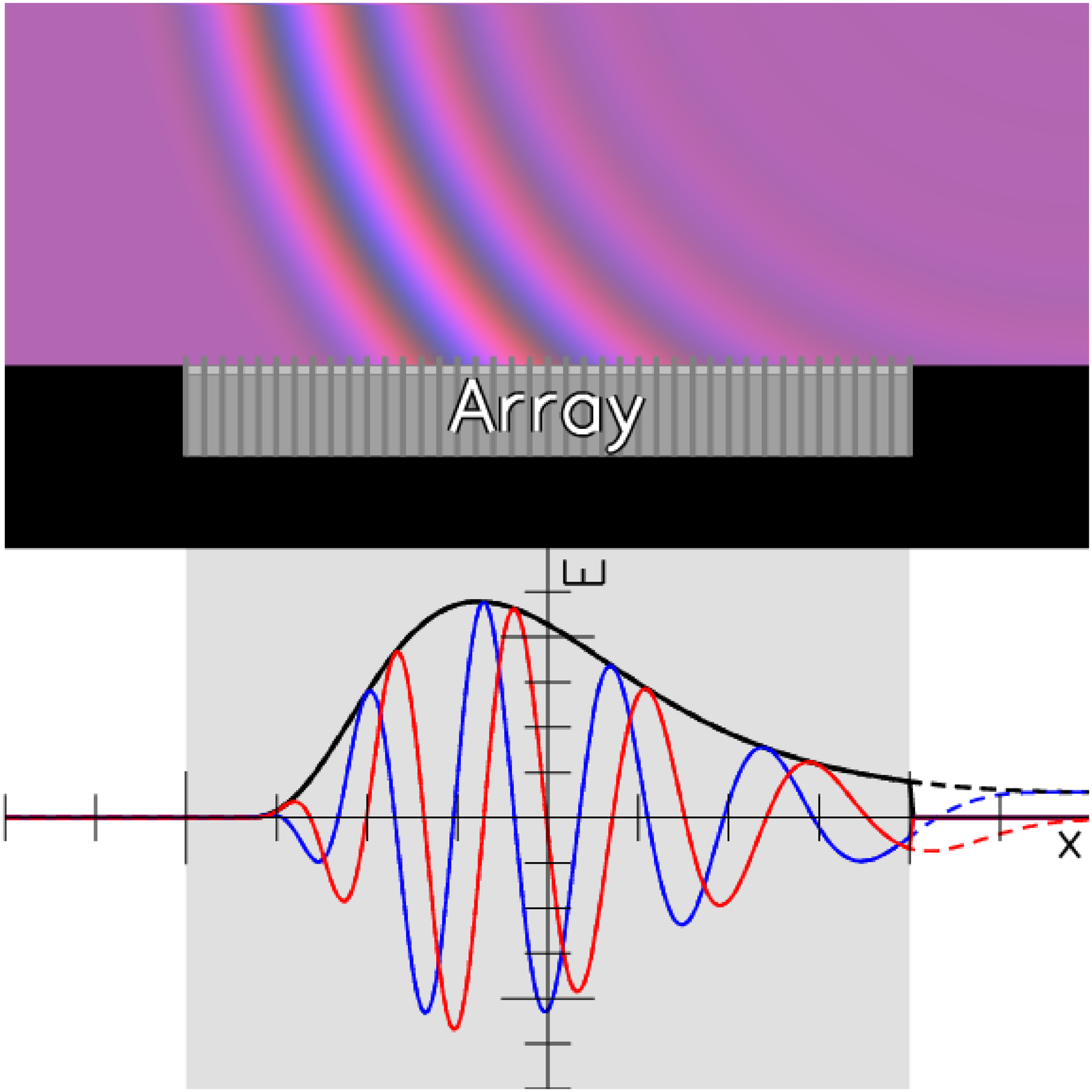}
\caption{Interferometry can be performed using any set of basis functions, not just Fourier modes.  A smaller zenith angle (left) stretches out the waveform in space compared to larger zenith angles (middle).  Sources that are close effect a ``chirp'' in the waveform, stretching it out more on one side of the array than others (right).  The lines styles are the same as in Figure~\ref{fig:Interferometry}. \label{fig:GenericInterferometry}}
\end{figure*}

The principles of a generic waveform interferometer are essentially the same as those for a regular Fourier interferometer (Section~\ref{sec:InterferometryIntro}).  The array on the ground measures a horizontal slice out of the electric field.  The measured waveform at a given time is some kind of projection of the waveform in time.  If the source is infinitely far away, the projection simply stretches out the waveform depending on its altitude.  The waveform is stretched out more for a source at high altitude (equation~\ref{eqn:tGenericInterferometer}) than for a source on the horizon (see Section~\ref{sec:InterferometryIntro} and Figure~\ref{fig:GenericInterferometry}).  By measuring the amount the waveform is stretched out on the ground, one can reconstruct where on the sky the source is.

When the source is not at infinity, the generic waveform interferometer is analogous to the fractional Fourier interferometer (Section~\ref{sec:FracInterferometer}).  The source's position on the sky has a measurable parallax from one end of the array to the other.  Thus, the stretch of the waveform on the ground varies from one position to another because of the quadratic dependence of the delay time on position (equation~\ref{eqn:tGenericInterferometer}).  On the side of the interferometer closer to the source, the waveform is stretched more than on the far side.  This is illustrated in Figure~\ref{fig:GenericInterferometry} (right panel).  By measuring the chirping of the waveform, one gets information on its distance.

As with the usual Fourier interferometer, the time dependence of the electric field signal allows one to distinguish whether its shape at a given moment is due to the radiation's temporal variation or is some kind of geometrical effect.  We can see from equation~\ref{eqn:tGenericInterferometer} that the geometry of the array merely shifts the delay time with position.  Otherwise, the time structure of the signal is not changed.

\subsection{Why the Fourier transform?}

We have greater freedom in interpreting the time structure of a radio signal than generally appreciated.  What is so special about the Fourier transform?  I believe the ubiquity of the Fourier transform arises from four reasons.  The first is that the Fourier transform is well studied, with plenty of ``fast'' algorithms existing.  There are widely available free packages that are tuned to compute the FFT quickly.  Yet fast algorithms exist for other transforms.  Fast algorithms now exist for all LCTs, if not quite as sophisticated as for Fourier transforms.  Additionally, the discrete wavelet transform has been known for more than 20 years.

Second, complex Fourier modes are defined as exponentials, which are eigenfunctions of standard derivatives and integrals.  This simplifies the analysis with standard calculus, as opposed to wavelets for example.  Yet there is no reason, in principle, why we could not use a calculus based on different operations.

Third, the Fourier modes are solutions to the wave equations as encountered in physics, like Maxwell's equations.  Dispersion, and the effective velocity of the signal, depends on pure frequency only.  Dispersive effects are particularly important for radio waves passing through interstellar plasma.  A mode with several Fourier frequencies disperses.  By using the Fourier transform, one can account for the effects of dispersion simply.  This is less of an issue at high frequencies.

Finally, and I believe most important, the astronomical sources currently being studied are \emph{slowly varying} -- at least compared to one cycle of a radio wave.  Radio sources have well-defined spectra that can be integrated over time.  Specifically, the power spectral density of radio signals from typical astronomical sources does not rapidly change in the Fourier domain.  As the Fourier transform is natural for long-lived sources, there has generally not been a need for another basis.  

But as the study of transient sources becomes more prevalent, there will be greater need for waveforms that are well-localized in time (as in Figure~\ref{fig:GenericInterferometry}).  Although the Fourier basis can be used for short pulses, the time structure of the pulse is ``hidden'' in the phase information.  The Fourier basis power spectral density of a pulse is constant at all frequencies, the same as for generic white noise.  In fact, the ANITA instrument for measuring radio pulses from ultra-high energy neutrinos already employs a kind of ``pulse interferometry'' for this reason \cite{Gorham09,RomeroWolf13}.

\subsection{FRB\lowercase{s} and the Fractional Fourier Transform}

A possible testing ground for generalized interferometers are FRBs.  FRBs are essentially chirps that last for a few milliseconds at each frequency; they appear as thin curves in ``waterfall'' plots of the usual time-frequency domain \cite{FRBs}.  Assuming an astronomical origin, the different arrival times of the FRBs at different frequencies is caused by dispersion in plasma, parameterized by the dispersion measure (DM):
\begin{equation}
\Delta t = 4.1\ \msec \left(\frac{\rm DM}{\pc\ \cm^{-3}}\right) \left[\left(\frac{\nu}{\GHz}\right)^2 - \left(\frac{\nu - \Delta\nu}{\GHz}\right)^2\right]^{-1}
\end{equation}
Bannister et al. (2011) \cite{Bannister11} noted that the delay time has a linear approximation because of Taylor's theorem:
\begin{equation}
\label{eqn:LinearDispersion}
\Delta t \approx -8.3\ \sec \left(\frac{\rm DM}{1000\ \pc\ \cm^{-3}}\right) \left(\frac{\nu}{\GHz}\right)^{-2} \left(\frac{\Delta \nu}{\nu}\right).
\end{equation}

As I discussed in Section~\ref{sec:FracInterferometer}, linear chirps are separable with a FrFT of the correct order.  We must put the chirp variable in dimensionless terms to use the FrFT in equation~\ref{eqn:FrFT}.  This can be done by using any reference frequency $\nu_0$ as the unit for the frequency and $1/\nu_0$ as the unit for time; then the chirping is $(1/\nu_0^2) (d\nu/dt)$.  For a burst that is smeared out by plasma dispersion, and using the linear approximation (equation~\ref{eqn:LinearDispersion}), the ``angle'' of the FrFT we want is
\begin{equation}
\label{eqn:AlphaFRB}
\alpha \approx \frac{\pi}{2} - 1.2 \times 10^{-10} \left(\frac{\nu}{\GHz}\right)^3 \left(\frac{\nu_0}{\GHz}\right)^{-2} \left(\frac{\rm DM}{1000\ \pc\ \cm^{-3}}\right)^{-1}.
\end{equation}
For a very small DM, $\alpha \to 0$ because the pulse is basically the same at all frequencies.  Then we want to examine the signal essentially in time domain, where it appears as an undispersed pulse.  For a very large DM, $\alpha \to \pi/2$ because the pulse appears to be a nearly pure tone, which very slowly slides in frequencies over time.  Then we want to examine the signal essentially in frequency domain, because the pulse is smeared out to a narrowband tone.

The difference between $\alpha$ and $\pi/2$ in equation~\ref{eqn:AlphaFRB} seems extremely small, but it is in fact measurable.  As with the fractional interferometer (Section~\ref{sec:FracInterferometerLimits}), the difference ${\cal F}_{\alpha}$ and ${\cal F}$ becomes measurable when the phase difference between a pure tone and the chirp is $\sim 1$ radian.  The signal must be sampled for a time $t \ga 1/\sqrt{d\nu/dt} \approx \nu^{-1} / \sqrt{|\alpha - \pi/2|}$:
\begin{equation}
t \ga 91\ \usec\ \left(\frac{\rm DM}{1000\ \pc\ \cm^{-3}}\right)^{1/2} \left(\frac{\nu}{\GHz}\right)^{-1/2}
\end{equation}
Since radio surveys last much longer than a few microseconds, a FrFT is actually quite useful.

\subsection{The chirpolator as an FrFT}
The problem of searching for chirps in signals for a sparse array has been considered before by Bannister et al. (2011) \cite{Bannister11}.  Their algorithm for finding chirps involves multiplying the complex voltages at one element with the complex conjugate voltage at a second element.  Since the chirp at one element is just a time-delayed copy of the chirp at another element, the product can be shown to be a pure tone.  A Fourier transform of the product finds the frequency of the tone, giving information on the chirp.  They call this method the ``chirpolator'', and note that it is the maximum likelihood estimator for chirps as found in \cite{Gershman01}.

Not surprisingly, this operation turns out to be the FrFT in disguise.  The kernel of a FrFT with $\alpha$ given by equation~\ref{eqn:AlphaFRB} is (c.f., equation~\ref{eqn:FrFT})
\begin{equation}
K_{\alpha} (t, t^{\prime}) = \sqrt{1 - \cot \alpha} \exp[i \pi \nu_0^2 ((t^{\prime 2} + t^2) \cot \alpha + 2 t^{\prime} t \csc \alpha)],
\end{equation}
using the $\nu_0$ to take care of units.  In this case $t^{\prime}$ has the units of time, and we can in fact interpret it as a time delay.  Thus, the kernel of the FrFT becomes
\begin{multline}
\label{eqn:KernelChirpolator}
K_{\alpha} (t, -\Delta t) = \sqrt{1 - \cot \alpha}\ \exp[-i \pi \dot{\nu} (t + \Delta t)^2] \\
\times \exp[2 \pi i \nu_0^2(\cot \alpha - \csc \alpha) t \Delta t].
\end{multline}
Given a chirp with frequency derivative $\dot{\nu}$, the FrFT multiplies it by a conjugated, time-delayed copy of itself, and then takes a Fourier transform of the result.  This is precisely what the chirpolator does, up to a multiplicative constant \cite{Bannister11}.  A similar algorithm by Bannister et al. (2011) is the chimageator, which involves taking the spatial Fourier transform of the voltage product.

The chirpolator and the FrFT formulations have different strengths.  With the chirpolator, it is not necessary to know the frequency slide $\dot{\nu}$ beforehand to get a reasonable estimate of the chirp properties.  One can apply the kernel in equation~\ref{eqn:KernelChirpolator} without knowing what $\alpha$ is, up to a frequency shift.  In addition, the antennas do not have to be in a rectangular grid to apply the chirpolator.  The FrFT's first advantage is that it is conceptually simpler -- just a rotation in Wigner space (Figure~\ref{fig:FrFTGeometry}).  Perhaps more importantly, the complexity of a Fast FrFT only goes as $\hat{N} \ln \hat{N}$ for $\hat{N}$ elements, whereas the chirpolator's complexity goes as $\hat{N}^2$.  One might consider combining these methods' advantages by using the chirpolator with a few elements to get a decent estimate of $\alpha$, and then using the FrFT on all of the complex voltages in the array to extract a precise characterization of a chirp with all of the data.

\section{Fully generalized interferometry}

\subsection{Interferometry in $L_2$ space}
All of the tools of functional analysis can generalize interferometry fully, as they have done for signal processing.  The electromagnetic field is a complex function of position.  Formally, we need four real numbers to describe it, one for each Stokes' parameter.  But for simplicity, just consider a single scalar, the electric field intensity at a given point as a function of time, $E(t)$.  ($E(t)$ can also stand for any other wave, including sound.)  Any realistic signal lasts only a finite time and contains a finite amount of energy:
\begin{equation}
\int_{-\infty}^{\infty} |E (t)|^2 dt < \infty
\end{equation}
But otherwise, $E^{\prime}$ may in principle be any square-integrable function.  

$E$ is therefore an element of the Hilbert space $L_2$, which represents all square-integrable functions.\footnote{The space $L_2$ is perhaps most familiar as the space of all quantum wavefunctions.}  It has an inner product defined by
\begin{equation}
\label{eqn:InnerProduct}
\InnerProduct{f}{g} = \int_{-\infty}^{\infty} \overline{f(t)} g(t) dx.
\end{equation}
We can define various orthonormal bases for $L_2$ from orthogonal functions with $\InnerProduct{f}{g} = 0$.  The vector of $E$ in $L_2$ in any basis is given by its projection onto each basis vector using the inner product.  The important part is that we can change between bases.

The standard basis for $E$ is the Dirac position basis, with orthonormal vectors representing each $\delta(t - a)$.  In this basis, the components of $E$ are just given by its value at each position, as can be seen by plugging $\delta(t - a)$ into equation~\ref{eqn:InnerProduct}.  We can say a trivial interferometer merely gives $E$.  The Fourier basis is also common, with basis vectors of $\exp(2\pi i \nu t)$.  The Fourier basis components of $E$ in $L_2$ is just the Fourier transform of the function.  A normal Fourier interferometer resolves $E$ into this basis, inferring where radiation is coming from.  These do not exhaust the possibilities.  All LCTs on $E$ -- including the FrFT, Laplace transform, and Fresnel transform -- perform a basis change on $L_2$, where the basis vectors are the integral kernel.  A major theme of this paper has been to explore when these other bases are appropriate.

The LCTs are just the beginning.  Other sets of orthogonal functions include polynomials, Bessel functions, wavelets \cite{Hubbard98}, and chirplets.  In the most general sense, an interferometer projects $E$ into any $L_2$ basis.  The only question is whether the bases have any useful meaning, and whether the transform is stable both forwards and in reverse (it is not in general; \cite{McWhirter78,Provencher82}).  

An interferometer array measures a cross section of the electric field along a line or plane.  If the source radiates isotropically, then the different antenna locations detect the same signal but sample it with different time delays.  The radiation from a source at infinity arrives as planar wavefronts with no curvature, in which case the time delay is exactly proportional to position.  If the source is closer, the time delay is roughly a second-order function of position.  Thus, the signal measured by the array at a given instant should be a stretched or chirped version of the signal in time.  An interferometer decomposes $E(t)$ into a set of basis functions, and then finds the inner product of these basis functions and the signal measured in space to determine the direction and distance of the source.

We use Fourier interferometers because the Fourier basis is the natural one for the time structure of the electric field, which in turn is due to the steadiness of typically-studied astronomical sources (section~\ref{sec:TimeDomain}).  The electric signals from a distant source have a simple representation in this basis, making it convenient for calculation.  It is analogous to reference frames in cosmology: there is only one reference frame in which the cosmic expansion is isotropic at any point, so we usually do calculations in this frame.  But the laws of physics are the same in all reference frames, and nothing stops us from working in any frame we wish.  Likewise, although Fourier interferometers are computationally useful, nothing stops us from doing interferometry in any other basis of $L_2$.

\subsection{Nonlinear and quantum interferometers}
I have focused on linear transformations of the $E$ signal, but nonlinear transformations are also possible.  For a monochromatic input wave, a nonlinear interferometer produces a warped output image.  This allows us to distort the Wigner space distribution of a signal in arbitrary ways.  This might be used to fit signals with quadratic dependence of frequency on position (cubic dependence of phase on position), including those from sources that are very nearby.  Likewise, one could also warp the time-frequency basis of a signal to fit quadratic chirps.  The main obstacle nonlinear interferometer, though, is the computational cost of calculating a relevant nonlinear transformation.  

Quantum optics is an increasingly important field.  Photon arrival times is already an integral part of intensity interferometry of thermal sources \cite{Foellmi09}.  Quantum non-demolition measurements may also prove useful in detecting structures below the diffraction limit \cite{Kellerer14}.  Just as classical optics can be emulated with a standard digital computer, it may be possible to emulate arbitrary quantum optical systems with a quantum computer.  The catch is that when the quantum computer produces an answer, the wavefunction collapses or decoheres.  Since a quantum state cannot be exactly copied, it is lost forever to us and cannot be recovered.  Thus, while we can run a classically recorded signal through any number of simulated classical optical systems simultaneously, we can simulate at most one quantum optical system.  The fundamental limitation is that we cannot determine which detector a photon passes through.  Equivalently, the phase of a single photon cannot be measured \cite{Burke69}.

\subsection{SETI and generalized interferometers}
To end on an extremely speculative note, an expanded consideration of the possible basis functions of a signal may have implications for the Search for ExtraTerrestrial Intelligence (SETI) \cite{Tarter01}.  So far, the most common assumption has been that a deliberate message would appear as an extremely narrowband signal \cite{Horowitz93}.  Frequency combs are another possibility \cite{Cohen95}, as are very short pulses \cite{Siemion10}.  Messerschmitt \cite{Messerschmitt14} argued that discrete points within blocks of time-frequency space could encode messages with optimal energy efficiency.

The Fourier basis seems natural to us, so we naturally think of tones and pulses.  But with SETI we have to extrapolate to entities with potentially very different psychologies and mathematics far beyond our comprehension.  For example, the exponential functions that make up the Fourier basis functions are the eigenfunctions of the standard differentiation and integration operations.  But are these the only possible operators that an ETI may base their calculus around?  If an ETI has studied quantum mechanics for a long time, one could imagine that the wavefunctions of the quantum harmonic oscillator may seem a more natural basis set.  Or consider that wavelets were essentially unknown to the physics community \cite{Hubbard98} when modern radio SETI was first proposed \cite{Cocconi59}.  

\begin{figure}
\includegraphics[width=8cm]{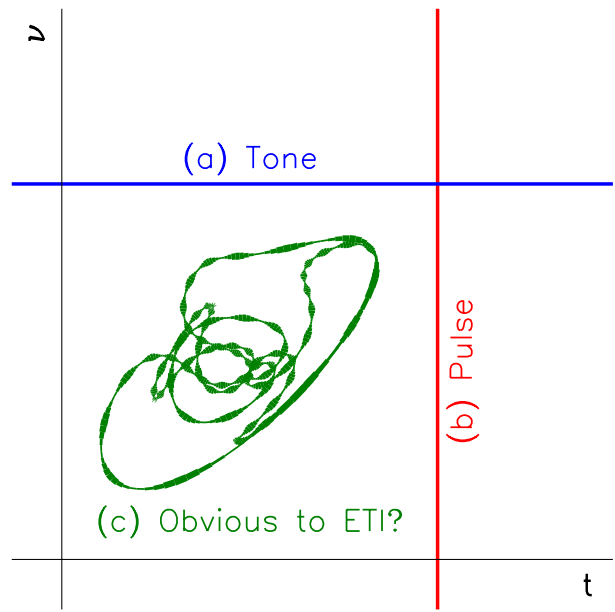}
\caption{The most commonly sought signal in radio SETI is a narrowband signal, approximating a tone (a).  Rapid pulses (b) are another sought ETI signal.  But if an ETI does not use the Fourier basis, the signal may look very complicated to us in time-frequency space (c). \label{fig:SETIWigner}}
\end{figure}

In time-frequency space, basis functions that seem ``obvious'' to a long-lived ETI may appear very complicated to us if they are not using the Fourier transform (Figure~\ref{fig:SETIWigner}).  A more general way to look for ETI signals is to search for strong signals that contain both fine-grained (or ``thin'') features and big, smooth regions in time-frequency space.  Narrowband signals, combs, pulses, and sets of discrete points, are just a few examples of such signals.  Leaving behind smooth 1D curves, we could imagine a signal that looks like a 2D pattern in time-frequency space, such as a Sierpinski triangle.  Known natural phenomena generally produce signals that are smeared out, with no fine-grained features, while white noise appears uncorrelated in all bases, with no big, smooth regions.  Even then, if an ETI uses some nonlinear (disconnected, even) mixture of time and frequency, the signals may just appear as blobs in time-frequency space.  Wavelets, after all, use a three-dimensional signal representation \cite{Hubbard98}.  In addition, looking for mixed fine-grained and smooth features in time-frequency space may be a strategy biased towards finding patterns that human perception emphasizes (such as edges in the visual field), which are contingent on human evolution.

Of course, it becomes impractical to search for \emph{all} possible signals, unless they are distinguished by some other way, like being extraordinarily bright signals from a Solar-type star.  Most of the possible signals look simply like white noise.  We may be guided by the assumption that the ETIs' resources are limited.  The LCTs and other transforms like the wavelet transform are advantageous in that they can be performed in only linear computational time.  In contrast, a typical linear transformation is a matrix multiplication that requires at least quadratic computational time. But ultimately, we may just have to follow the example of Bell Burnell in discovering pulsars \cite{BellBurnell77}, and examine time-frequency space (or any other obvious representation that occurs to us) closely for anything out of the ordinary \cite[c.f.,][]{Tarter01}.

\section{Conclusions}
There is nothing inherently special about the Fourier basis that requires its use for interferometry.  In time domain, the signal can be decomposed into any set of basis functions.  Basis functions that are more wideband but localized in time may be more useful when characterizing very short transients.  The position dependence of the electric field measured by a planar array is related to the signal in time domain for an isotropic source, which is stretched out due to projection effects if the source is away from the zenith, and is chirped if the source is nearby.  The Fourier modes are natural mainly because astrophysical sources vary slowly in time.

Digital interferometers simulate optical systems.  The standard Fourier transform performed by interferometers focuses light coming from infinity.  I have described in this paper fractional interferometers, which use the Fractional Fourier transform.  These focus light from sources that are closer than infinity.  I have demonstrated that there are many sources that are in the regime where a fractional interferometer is useful, including atmospheric radio transients, artificial satellites, and VLBI observations of objects within the Solar System. The FrFT with orders very close to 1 can be implemented by combining the standard Fourier transform and chirp multiplications.  I have shown that the fractional interferometer is analogous to the chirpolator \cite{Bannister11}.  Other Linear Canonical Transforms simulate different optical elements.

Since interferometers these days record signals digitally, it is possible to simulate optics that are not physically possible.  I considered the possibility of digitally emulating negative refractive index superlenses.  These would be useful for constructing images of sources that are within one wavelength of the array, which may be useful at very low frequencies.  These techniques cannot be applied to sources that are astronomically distant, because noise overwhelms any remaining signal from evanescent waves.  This is related to the impracticality of using the Laplace transform as a basis for interferometers -- the separation of exponentials is generally an ill-posed problem \cite{Istratov99}.

My discussion in this paper has been general and qualitative.  There are many practicalities that need to be considered in practice.  Most importantly, it is important to extend the results to the sparse arrays typical of interferometers, instead of the dense arrays used in FFTTs.  What is required is a generalized version of visibilities and the van Cittert-Zernike theorem.  Secondly, I have ignored the sky's curvature.  To accurately implement wide-field general interferometers, fractional and other extensions of spherical harmonics \cite{Shaw14} would be more appropriate, or the sky could be divided into small facets \cite{Tegmark10}.  In addition, I have ignored the polarization of the electric field.  Calculations of the signal-to-noise ratio of observations by general interferometers in the presence of noise must be carried out.

\begin{acknowledgments}
I acknowledge support from the Institute for Advanced Study.
\end{acknowledgments}

\end{document}